\documentclass{article}[12pt]

\usepackage{setspace}
\usepackage{url}
\usepackage[pdftex]{graphicx}
\usepackage{amsmath,amssymb}
\usepackage[numbers]{natbib}
\title{Intermittent Control Properties of Car Following: \\Theory and Driving Simulator Experiments}
\author{\\Ihor Lubashevsky and Hiromasa Ando\\[2em]
	University of Aizu}

\begin{document}
\thispagestyle{empty}
\maketitle	

\tableofcontents

\begin{abstract}

A rather simple car driving simulator was created based on the available open source engine TORCS and used to analyze the basic features of human behavior in car driving within the car-following setups. Eight subjects with different skill in driving real cars participated in these experiments. They were instructed to drive a virtual car without overtaking the lead car driven by computer at a fixed speed and not to lose sight of it. Moreover, these experiments were conducted with four different speed including 60km/h, 80km/h, 100km/h, and 120km/h. Based on the collected data the distribution of the headway, velocity, acceleration, and jerk are constructed and compared with available experimental data collected previously by the analysis of the real traffic flow. A new model for car-following is proposed capture the found properties. As the main results we draw a conclusion that the human behavior in car driving should be categorized as a generalized intermittent control with noise-driven activation of the active phase. Besides, we hypothesize that  the extended phase space required for modeling human actions in car driving has to comprise four phase variables, namely, the headway distance, the velocity of car, its acceleration, and the car jerk, i.e., the time derivative of the car acceleration. This time, the time pattern of pedal pushing and the distribution of time derivative of pedal was utilized in addition to previous variables. Moreover, all subjects’ driving data were categorized as some styles with their shapes.
\end{abstract}

\section{Introduction}

According to the modern point of view  \cite{gawthrop2011intermittent,loram2011human,balasubramaniam2013control,milton2013intermittent,asai2013learning} human control over unstable mechanical systems should be categorized as intermittent. As far as human behavior is concerned, intermittency implies discontinuous control, which repeatedly switches off and on instead of being always active throughout the process. As a result, the actions of a human operator in controlling a mechanical system form a sequence of alternate fragments of phases of his passive and active behavior. According to the current state of the art, this type control being rather efficient on its own is a natural consequence of human physiology (see., e.g., \cite{loram2011human}). 

The concept of event-driven intermittency is one of the most promising approaches to describing human control. It posits that the control is activated when the discrepancy between the goal and the actual system state exceeds a certain threshold. Models based on the notion of threshold can explain many features of the experimentally observed dynamics \cite{gawthrop2011intermittent,milton2013intermittent}. However, much still remains unclear even in the case of relatively simple control tasks, such as real \cite{balasubramaniam2013control,cabrera2002onoff,milton2009balancing} or virtual \cite{loram2011human,foo2000functional,bormann2004visuomotor,loram2009visual} stick balancing. For instance, the mechanism behind extreme fluctuations of the systems under human control (resulting, e.g. in stick falls) still has to be explained \cite{cabrera2012stick}. 

Recently, a novel concept of noise-driven control activation has been proposed as a more advanced alternative to the conventional threshold-driven activation \cite{zgonnikov2014Inerface}. It argues that the control activation in humans may be not threshold-driven, but instead intrinsically stochastic, noise-driven, and stems from stochastic interplay between operator’s need to keep the controlled system near the goal state, on the one hand, and the tendency to postpone interrupting the system dynamics, on the other hand. To justify the noise-driven activation concept  a novel experimental paradigm: balancing an overdamped inverted pendulum was employed~\cite{zgonnikov2014Inerface}. The overdamping eliminates the effects of inertia and, therefore, reduces the dimensionality of the system. Arguably, the fundamental properties and mechanisms of human control are more likely to clearly manifest themselves in such simplified set-ups rather than in more complicated conventional experimental paradigms. In the frameworks of human intermittent control with noise-driven action the transition from passive to active phases is considered to be probabilistic, which reflects human perception uncertainty and fuzzy evaluation of the current system state before making decision concerning the necessity of correcting the system dynamics.

Driving a car in following a lead car is a characteristic example of human control, which allows us to suppose that the intermittency of human control should be pronounced in the driver behavior and affect the motion dynamics essentially. The general objective of our study is to elucidate how the basic properties of human control manifest themselves in the characteristics of car driving. Another reason of choosing the given subject is to understand which factors are responsible for the found characteristic properties of real traffic flow, at
least, some of them. There could be two different in nature factors reflecting in them. The first one is various heterogeneities of road structures. The second one is basic human properties noted above. The use of car driving simulators could enable us to separate their
contributions because we control the virtual environment completely.

\section{Car Driving Simulator}

In the conducted experiments we explored a car-driving simulator created based on the open source engine TORCS (The Open Racing Car Simulator). It is a highly portable multi platform car racing engine widely used in ordinary and AI car racing games as well as a research platform \cite{TORCS_OffisSite2014}.	A commercially available high-precision steering wheel with pedal set (Logitech G27 Racing Wheel) was used in the experiments.

\begin{figure}[t]
	\begin{center}
	  	\includegraphics[width=0.8\columnwidth]{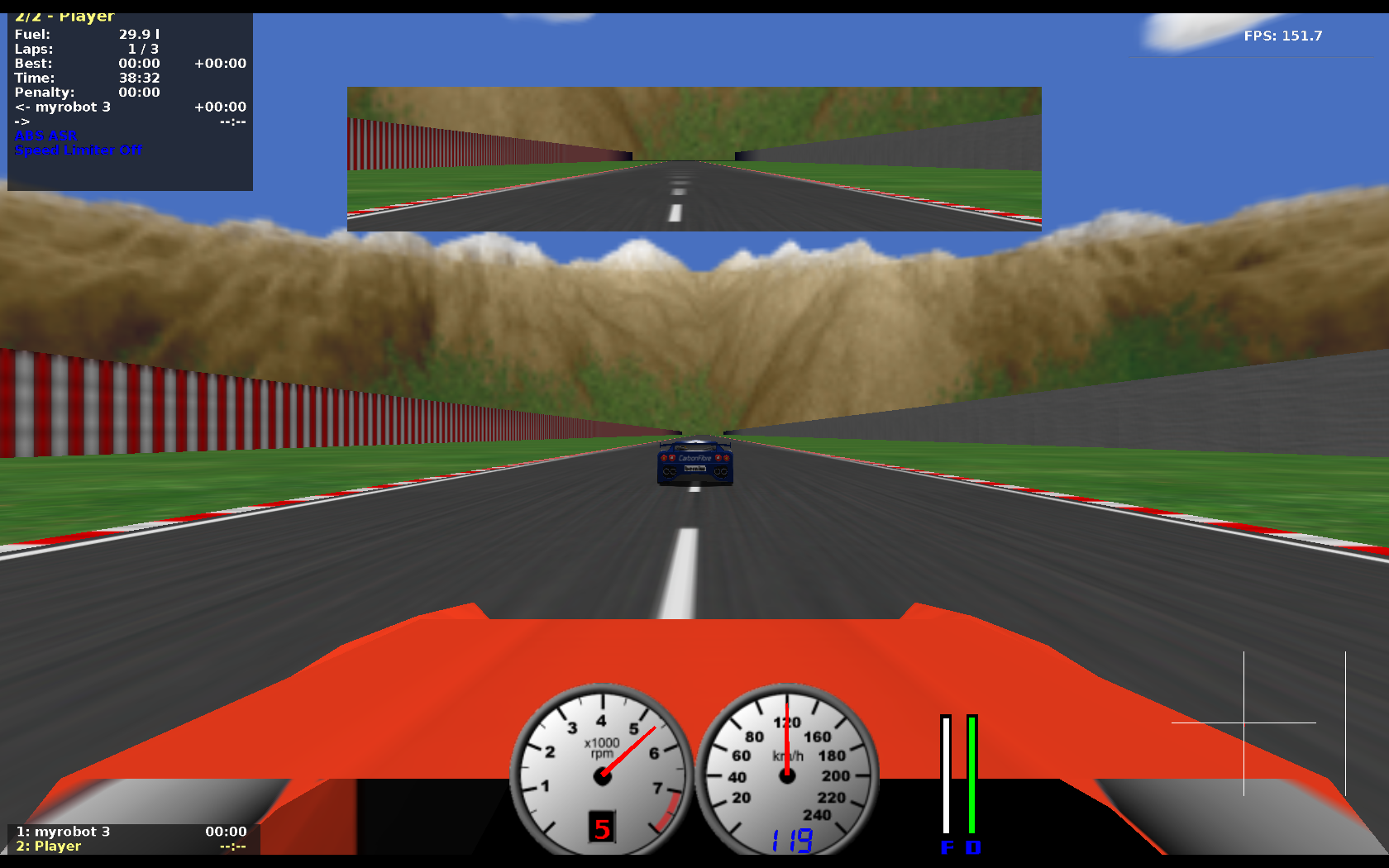}
	\end{center}
 	\caption{Screenshot of car-following experiments. The car ahead (in blue) is driven by computer at a fixed speed.}
 	\label{fig:one}
\end{figure}

The experiments were implemented within the car-following setup, i.e., the subjects were instructed (\textit{i}) to follow the lead car without overtaking it and (\textit{ii}) to keep a certain headway distance not to lose sight of the lead car. The other details of driving their cars subjects  may chose according to their individual preference. A screenshot in Fig.~\ref{fig:one} illustrates the car-following experiments. 

\begin{figure}[p]
	\begin{center}
		\includegraphics[width = \columnwidth]{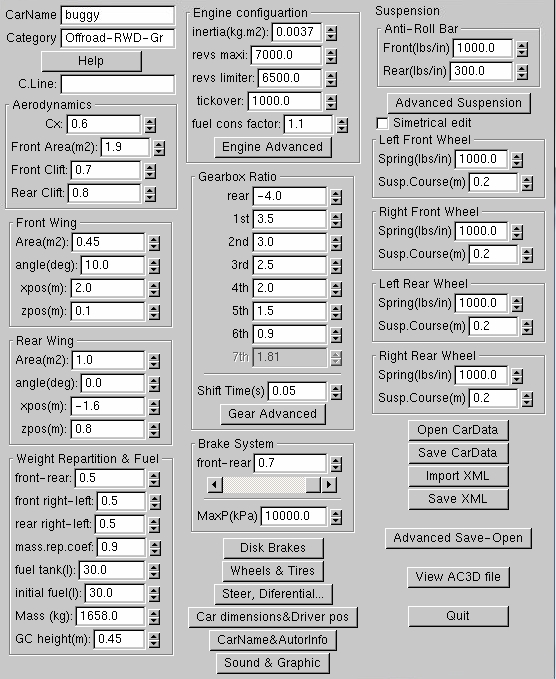}
	\end{center}
	\caption{Details of the car model driven by subjects. Screenshot of the interface of ``torcs-car-setup-editor'' v0.11 by Vicente  Mart\'{\i} Centelles \cite{tceditor}.}
	\label{fig:four_1}
\end{figure}

\begin{figure}[p]
	\begin{center}
		\includegraphics[width = \columnwidth]{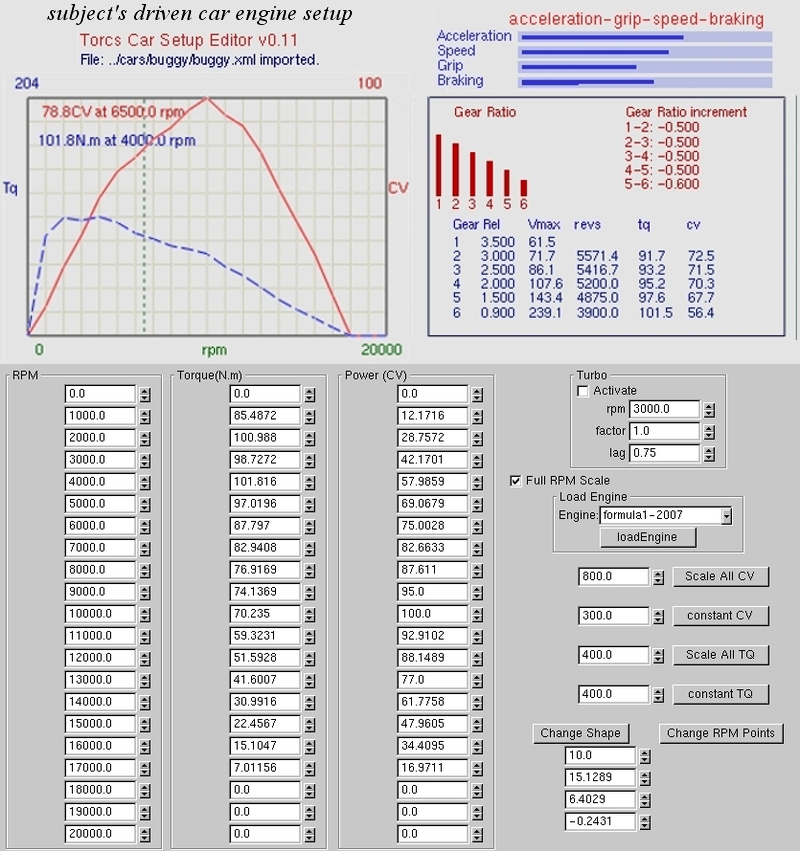}
	\end{center}
	\caption{Details of the engine setup for the car driven by subjects. Screenshot of the interface of ``torcs-car-setup-editor'' v0.11 by Vicente  Mart\'{\i} Centelles \cite{tceditor}.}
	\label{fig:four_2}
\end{figure}

\begin{figure}[p]
	\begin{center}
		\includegraphics[width = 0.8\columnwidth]{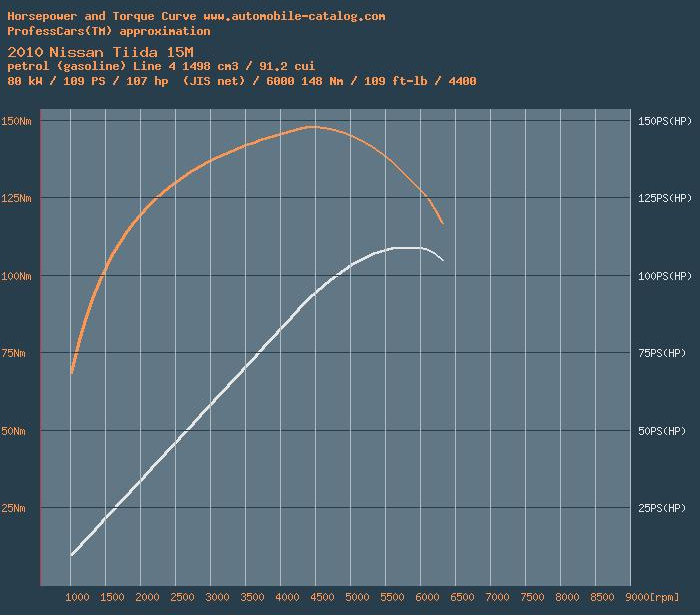}
	\end{center}
	\caption{The horsepower and torque characteristics of real cars, Nissan Tiida is chosen as an example, the plot has been generated by  ProfessCars$^\text{TM}$ software based on the factory data, \protect\url{http://www.automobile-catalog.com}\,.}
	\label{fig:tiida}
		\begin{center}
			\includegraphics[width=0.47\columnwidth]{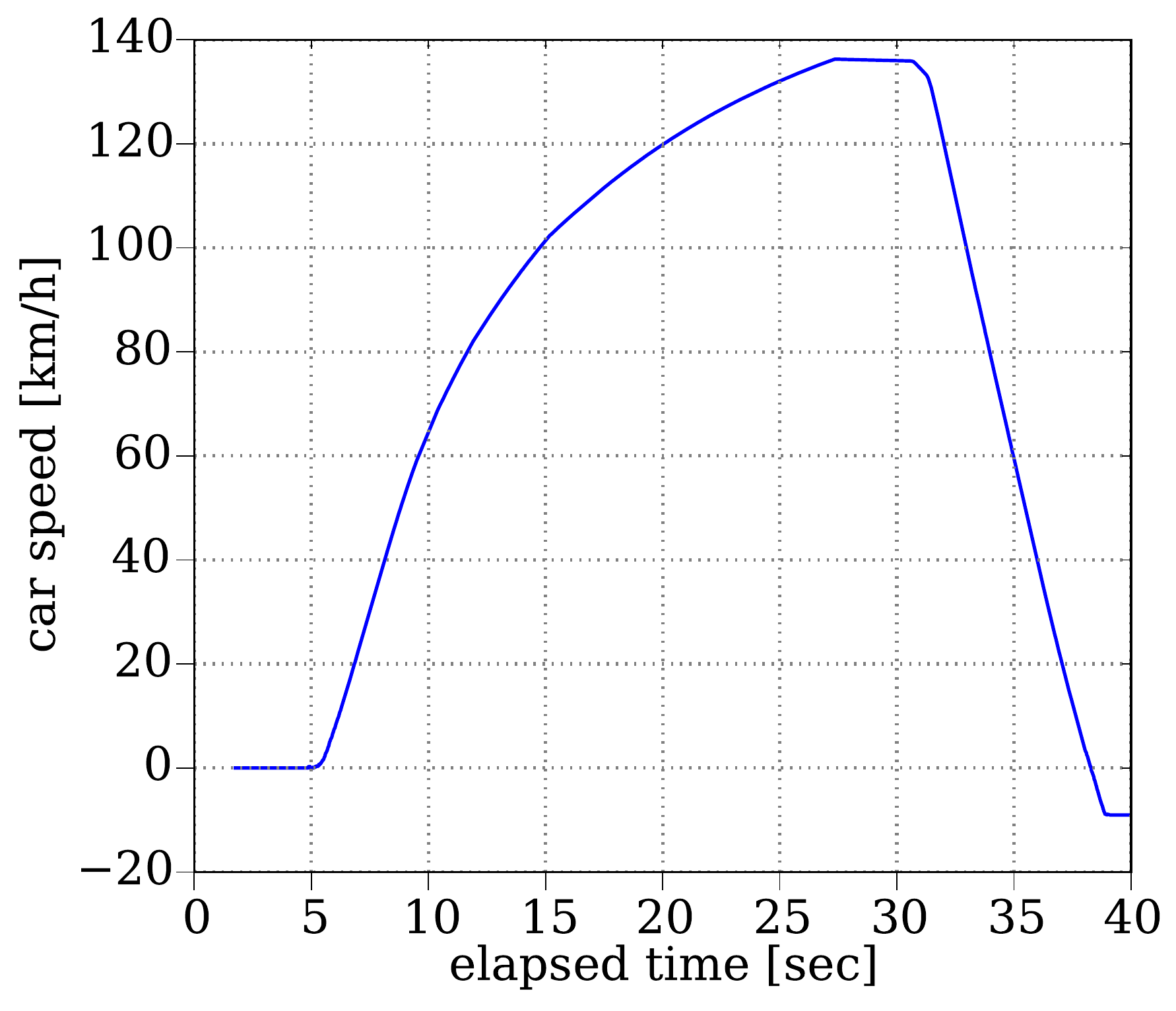}\quad
			\includegraphics[width=0.47\columnwidth]{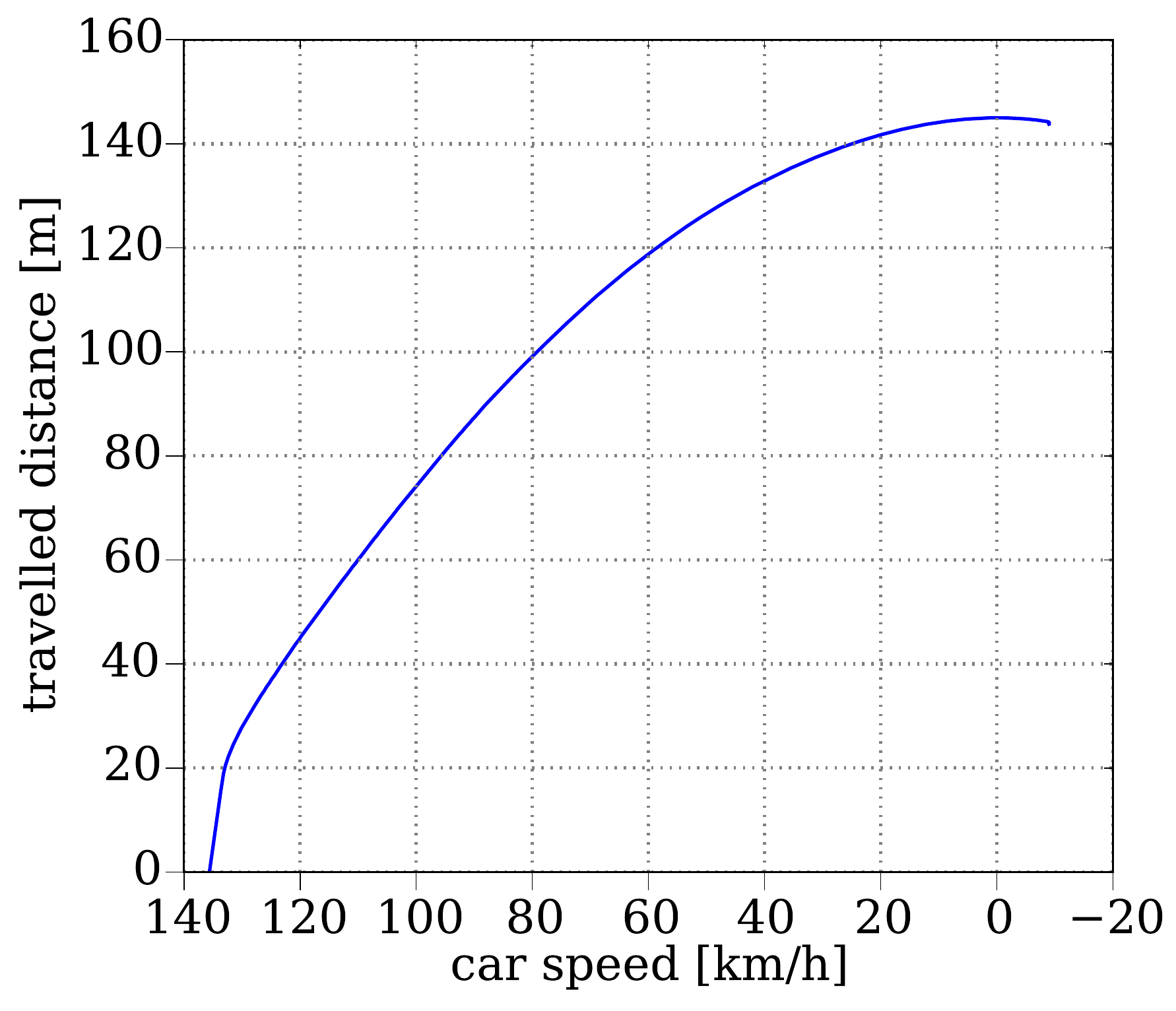}
		\end{center}
		\caption{The dynamic characteristics of the car model driven by subjects. The maximal speed growth (left plot) and the dynamics of the extremal brake (left and right plots) when the acceleration and brake pedals are pressed up to their limit positions. The last part of the speed time pattern corresponds to negative values because the braking pedal of the steering unit induces backward motion with a relatively low speed.}
		\label{fig:acdc}
\end{figure}

A car model driven by subjects was created using the software ``torcs-car-setup-editor'' (v0.11) \cite{tceditor}. The characteristics of its engine and the other mechanical parameters of this car model were chosen to imitate the features of real vehicles belonging to the subcompact and compact car classes, in particular, Nissan Tiida (15M, 2010) may be treated as an example of intermediate models between the cars of the two classes. Figures~\ref{fig:four_1} and \ref{fig:four_2} depict the details of the employed car model including the dependence of the engine power and torque curves on rmp (rotations per minute). Figure~\ref{fig:tiida} illustrates the engine horsepower and torque curves for a real car.

The car dynamic characteristics of maximal acceleration and brake, i.e., the speed growth and the brake process when the acceleration and brake pedals of the steering unit are pressed up to the limit positions are illustrated in Fig.~\ref{fig:acdc}.  As seen, under such conditions the car model gets the speed of 100~km/h from the start by 10~sec which matches the technical characteristics of real vehicles of the considered car classes, e.g., for  Nissan Tiida this time is about 11--12 sec \cite{tiida2004}. The braking distance, i.e., the distance a vehicle will travel from the point when its brakes are fully applied to when it comes to a complete stop, is evaluated as approximately 140~m for the initial speed about 140~km/h, which also corresponds to the real data, depending on the road conditions the break distance can change from 100--140~m for the given speed \cite{brakecalcul2016}.



\subsubsection*{Track details}

For these experiments a special track was constructed. It has a rectangular form (with smoothed corners) whose longest straight parts are of length about 70~km. One trial of experiments is implemented via driving along one of the longest parts,  which enabled us not to take into account the effects of road curvature. The width of the track road is 15.00 meters.  The roadside of the track includes a special pattern of stripes enabling a subject to get feeling of the current speed of the driven car (Fig.~\ref{fig:one}).  
 
\subsubsection*{Recorded data} 

The data recorded at frequency of 50~Hz (with time interval about 20~ms) are saved in a text-file. They include the individual information about the lead and following cars with respect to their position on the track, the velocity, and the degree of pressing the acceleration and brake pedals. The car acceleration and the rate of pedal position variations were calculated using the 1D symmetric Savitzky–Golay filter with the fitting polynomial of order $n=3$ and the window length of 25 points. As verified directly, this size is optimal with respect denoising the recorded data and saving the details of subject's actions. Besides, the recorded data of car position generated by the car-driving simulator are not of the accuracy required for \textit{calculating} the car speed, so the car acceleration and jerk were calculated based on the speed data. The Python components NumPy (ver. 1.10) and Matplotlib (ver. 1.5.1) were used for numerical manipulations and visualization under Python 3.4.

\section{Experiment Setups}

The set of car-following experiments consisted of trials when the lead car speed was set equal to $V = 60$~km/h, 80~km/h, 100~km/h, and 120~km/h. Each of these trials was continued for 60 minutes totally with possible breaks caused, at least, by the necessity to move from one long-distant track to the the other. The subjects were allowed to brake a trial for other reasons, for example, to have a short-time rest;  almost all the subjects divided their driving experiments into a few parts of duration of 20--40~min. 

\begin{table}[t]
	\caption{Subjects' characteristics and their driving experience.}\label{Tab1}
	\begin{center}
		\begin{tabular} {|c|l|l|l|} \hline
			ID  & Driving License; issued & Driving Frequency \\ \hline
			1   & Yes;  2 months ago & Rarely \\ \hline
			2   & Yes;  3 years ago & Rarely \\ \hline
			3   & Yes;  1 years ago & Sometimes \\ \hline
			4   & Yes;  3 years ago & Often \\ \hline
			5   & Yes;  4 years ago & Rarely \\ \hline
			6   & Yes;  2 years ago & Daily \\ \hline
			7   & Yes;  4 years ago & Rarely \\ \hline
			8   & Yes;  3 years ago & Daily \\ \hline
		\end{tabular}
	\end{center}
\end{table}


Eight male students of age around 22-25 participated in the experiments and each of them was asked to do it twice; Table~\ref{Tab1} presents subjects' information.

\section{Results and Discussion}

The pedal position is the quantity controlled directly by drivers, the time variation in the other quantities---the car position, velocity, acceleration, etc.---are determined by the car mechanics and the pedal positions. In this sense they are controlled by drivers in an implicit manner. Therefore we put forward the idea of analyzing subject's actions studying, at the first step, the corresponding time patterns of the pedal positions and their time derivatives. There are at least to reasons for including the pedal position time derivative into the list of the main characteristics of driver actions. First, turning to our causal experience of car-driving it is clear that drivers focus their attention on the car arrangement in traffic flow, the current velocity and acceleration rather than on the particular position of the pressed pedal. However, after making decision on correcting the current state of car motion a driver consciously (or automatically based on the gained experience) slow or fast pushes or releases the corresponding pedal. In other words, the rate of pedal movement can be a quantity controlled by the driver's active behavior. Second, the main quantity that is controlled directly by a human operator governing the dynamics of a given system and, in its turn, quantifies the operative actions in the active phase of his intermittent control has to exhibit a special characteristic property. Namely its distribution function (i.e. histogram) has to possess a sharp peak at the origin \cite{zgonnikov2014Inerface}, which means that during the passive phase where the operator control is suspended this quantity does not change. In accordance with the obtained results, the pedal position time derivative is this type variable.              

\begin{figure}[t]
	\begin{center}
	  	\includegraphics[width=0.49\textwidth]{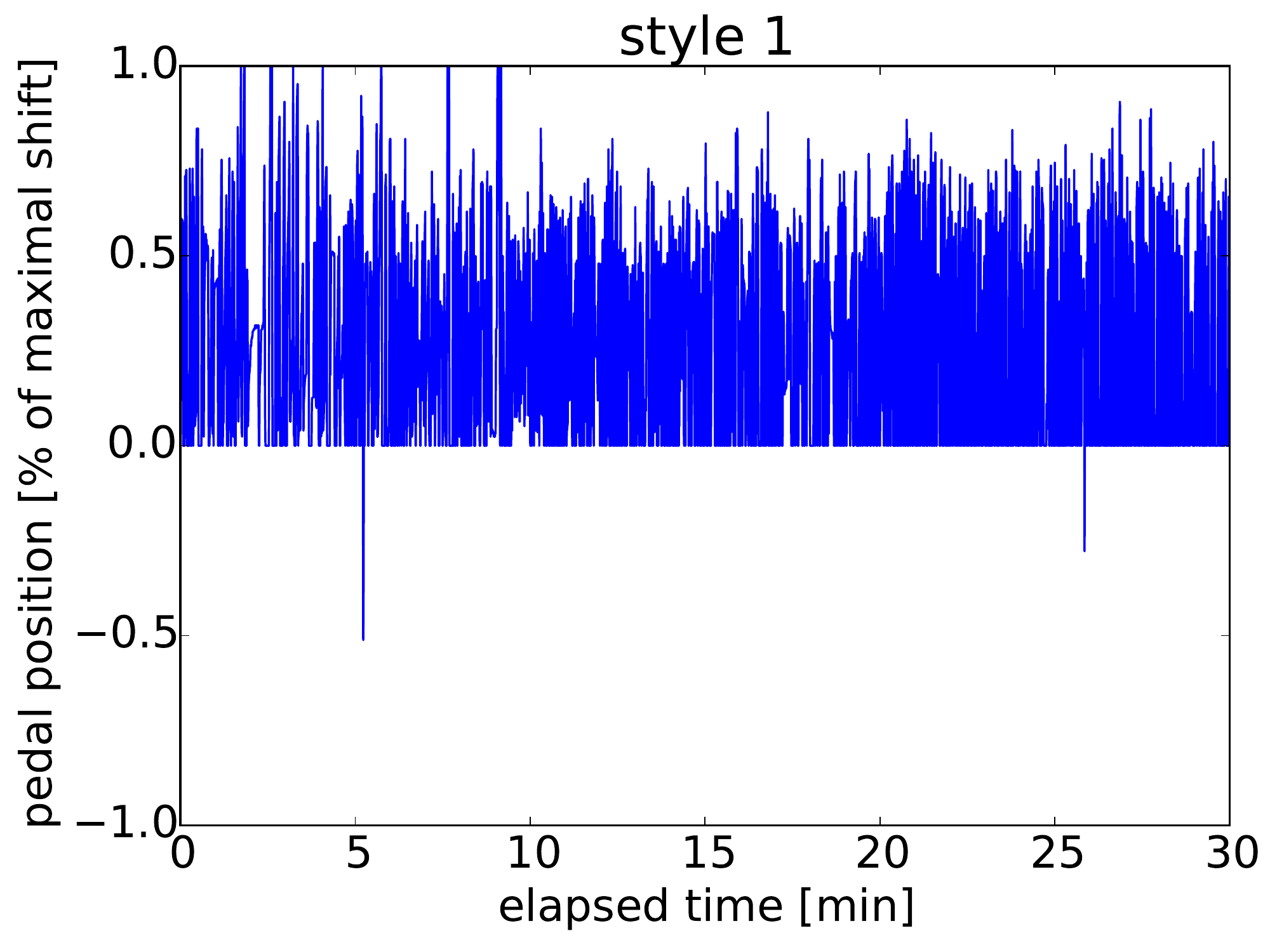}
        \includegraphics[width=0.49\textwidth]{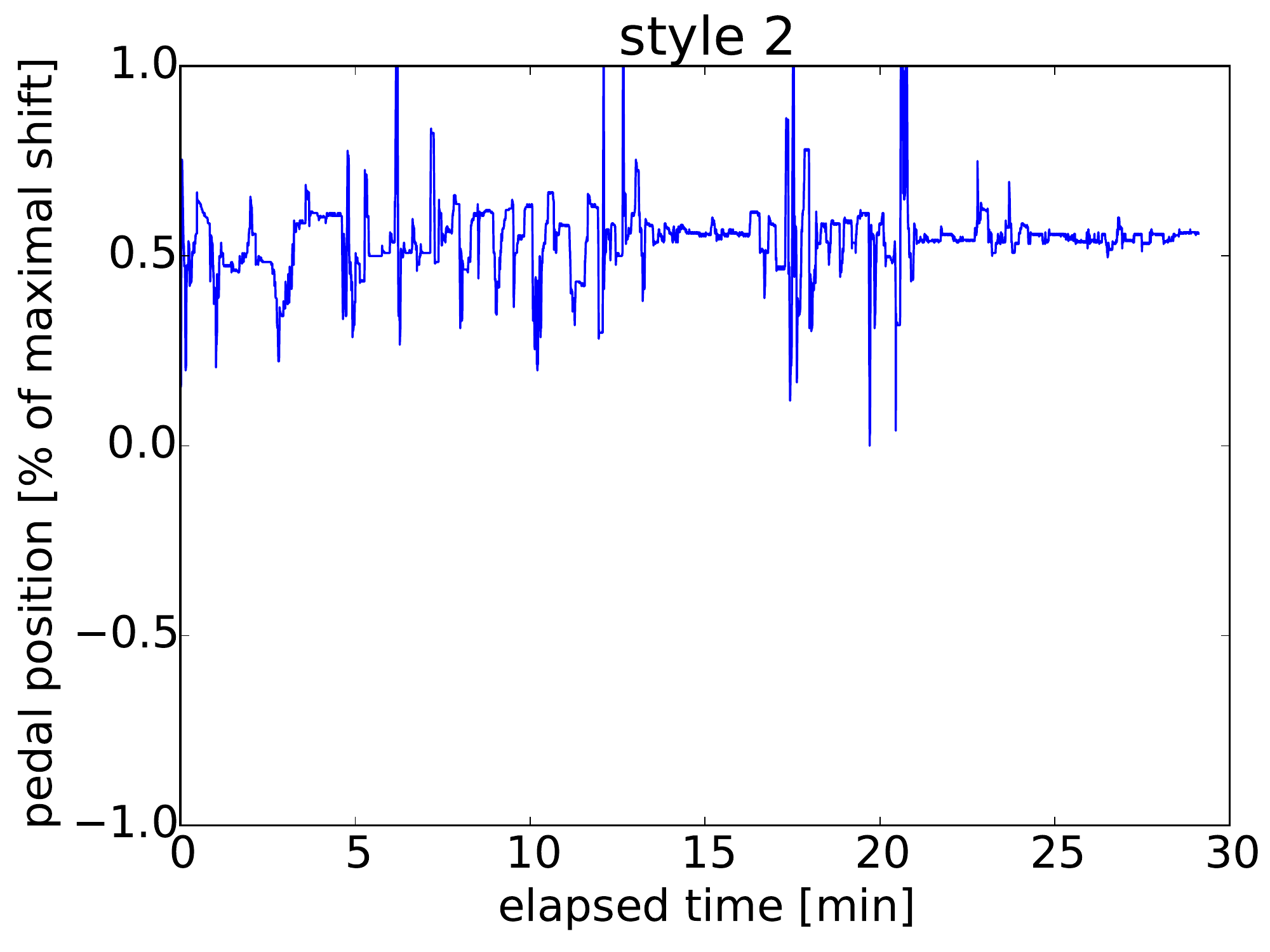}
		\includegraphics[width=0.49\textwidth]{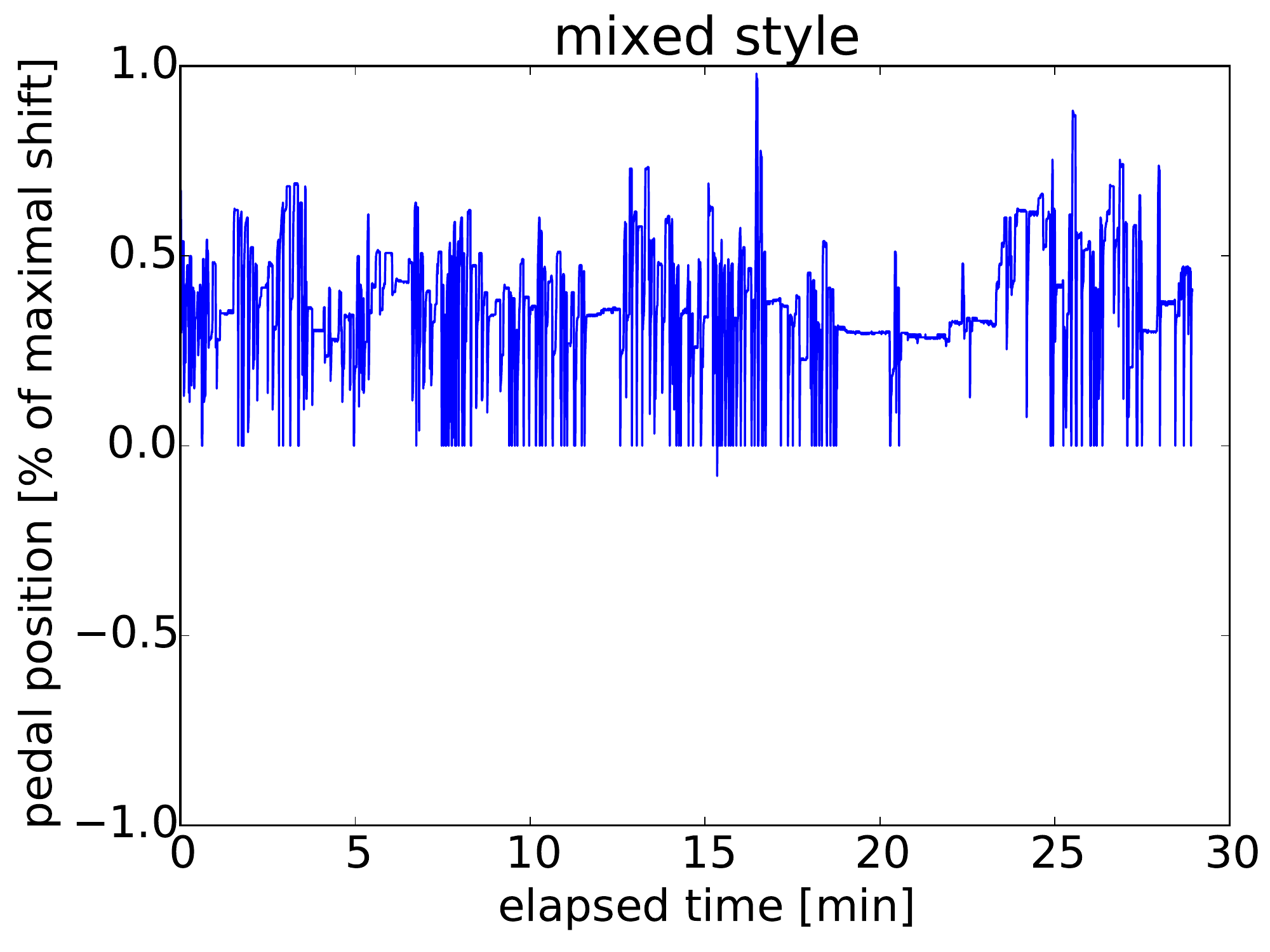}
	\end{center}
 	\caption{Characteristic forms of the pedal position time patterns being the reference points in classification of driving styles. The upper left plot visualizes the actions of subject 2 when the lead car velocity is $V = 80$~km/h, the upper right plot does it for subject 7 and $V=120$~km/h and the lower middle plot corresponds to subject 6 and $V=120$~km/h.}
 	\label{fig:style1_PT}
\end{figure}

Figure~\ref{fig:style1_PT} shows typical forms of the pedal position time patterns found in the conducted experiments. The time pattern classified as style~1 demonstrates the strategy of car driving when a driver pushes the acceleration pedal for a relatively short time then release it also for a short time and so on. This style enables a driver to keep up the desired velocity and headway without the precise control over the pedal position just changing the duration of pressing or releasing the pedal. The time pattern classified as style~2 demonstrates the opposite strategy of car-driving, when a driver is able to keep up the required pedal position for a relatively long time interval. The shown third style may be classified as a certain mixture of the first two basic styles. The present work has been confined to studying the characteristic properties of the two basic styles. The mixed style is worthy of individual investigation.

\begin{figure}[t]
	\begin{center}
		style 1\hspace{0.40\textwidth}style 2\\
		\includegraphics[width=0.49\textwidth]{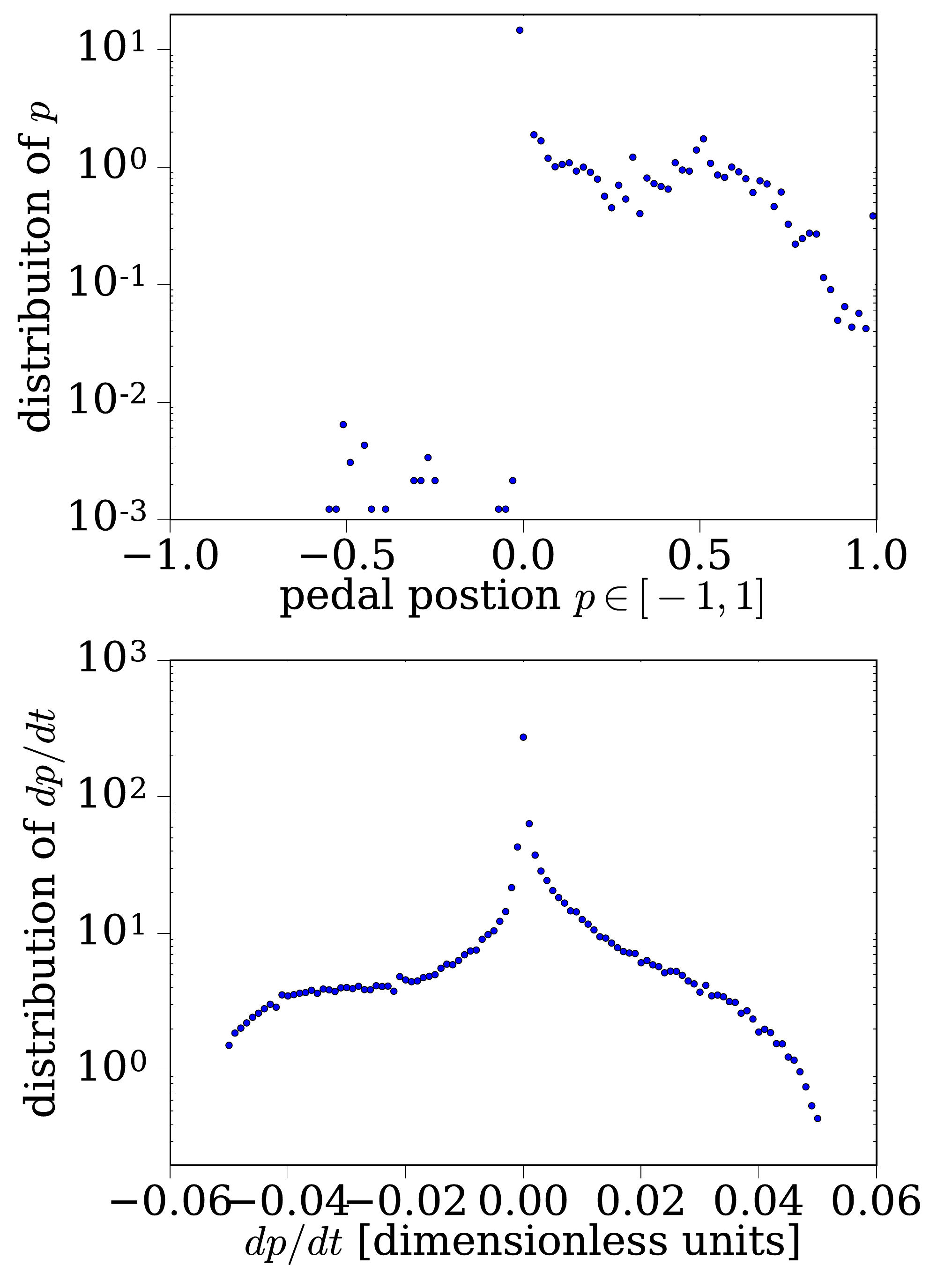}
		\includegraphics[width=0.49\textwidth]{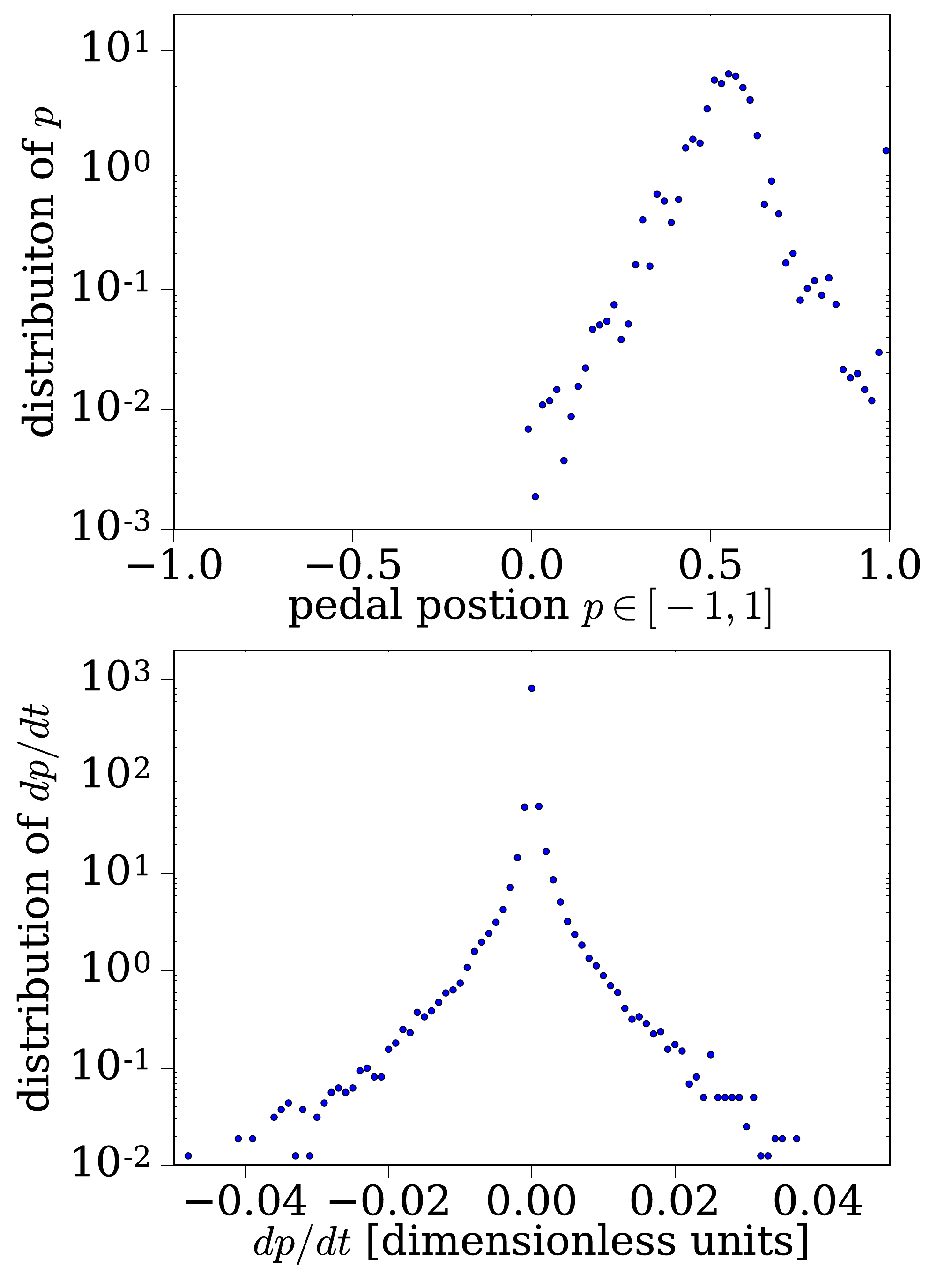}
	\end{center}
	\caption{Characteristic forms of the distributions of the pedal position $p$ and its time derivative $dp/dt$. Styles 1 and 2 are exemplified using the data for  subject 2 with $V = 80$~km/h and subject 7 with $V=120$~km/h.}
	\label{fig:pdp_styles12}
\end{figure}

Figure~\ref{fig:pdp_styles12} shows the typical form of the distribution of the pedal position and its time derivative for styles 1 and 2. As seen, the distribution of the pedal position, the $p$-distribution, are completely different, as should be expected. For style 1 the $p$-values are scattered rather widely in the possible acceleration interval (0,1) with a non-pronounced maximum at the required value $p_V$ for the steady-state motion. For style 2 the $p$-distribution is located near the corresponding value $p_V$ and has the form of the Laplace distribution. The distributions of the pedal position time derivatives, the $dp/dt$-distributions, differ for styles 1 and 2 in the form and scales. Nevertheless, all the found $dp/dt$-distributions possess a common feature, it is a sharp peak at the origin. We relate the appearance of this peak to the basic properties of human intermittent control.

\begin{figure}[p]
	\begin{center}
		style 1\hspace{0.40\textwidth}style 2\\
		\includegraphics[width=0.49\textwidth]{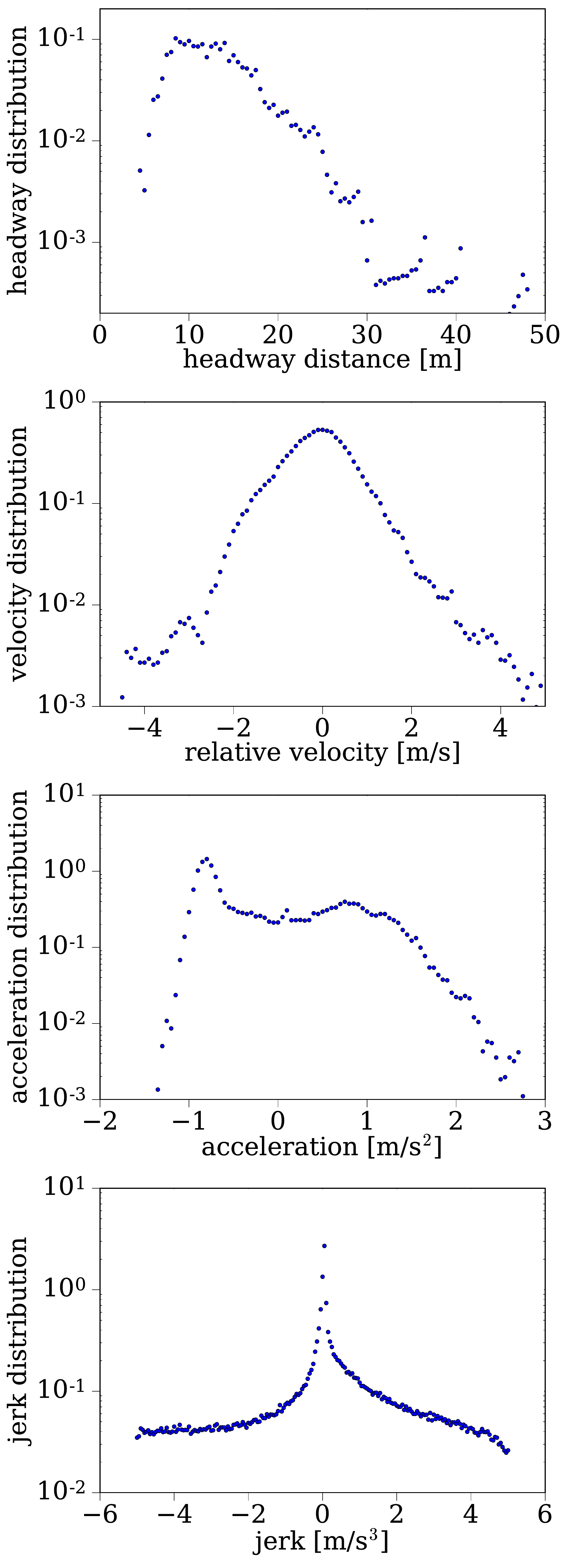}
		\includegraphics[width=0.49\textwidth]{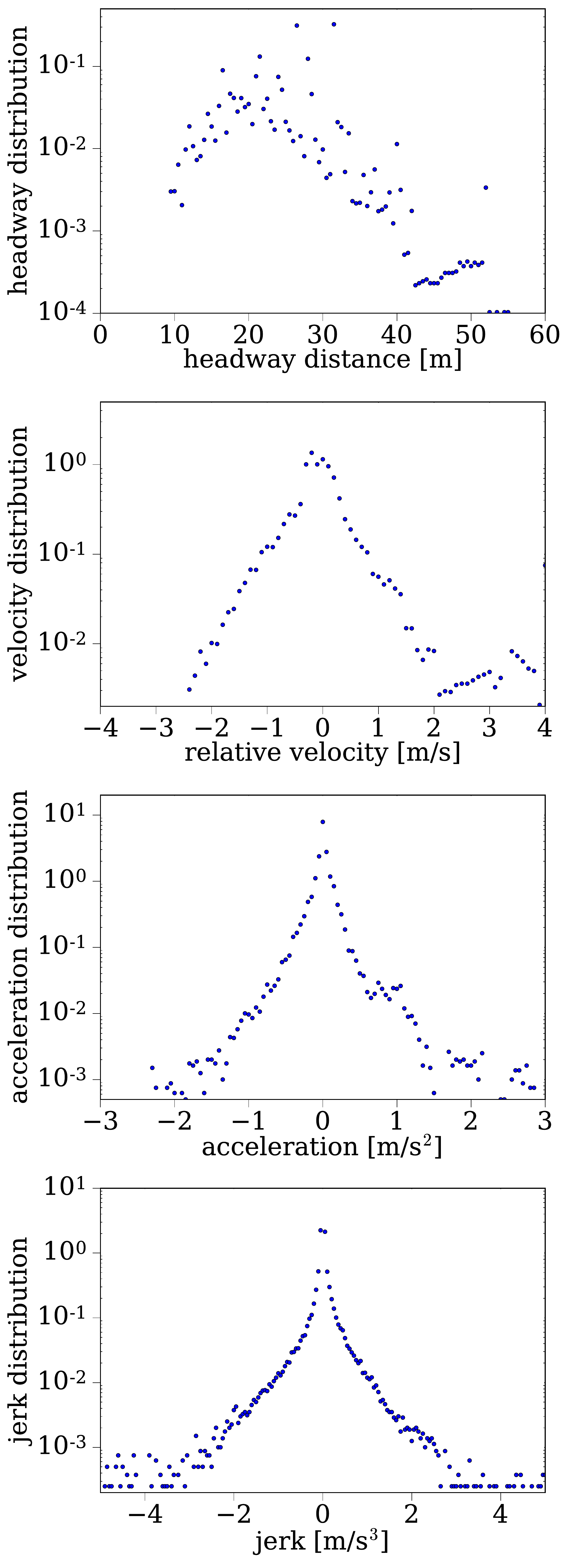}
	\end{center}
	\caption{Characteristic forms of the distributions of the headway distance, the relative velocity, the car acceleration and jerk. Styles 1 and 2 are exemplified using the data for  subject 2 with $V = 80$~km/h and subject 7 with $V=120$~km/h.}
	\label{fig:hvaj_styles12}
\end{figure}

The statistical properties of the other characteristics of the car dynamics, namely, the headway distance between the cars, their relative velocity, the following car acceleration, and the jerk, i.e., the time derivative of the car acceleration are shown in Fig.~\ref{fig:hvaj_styles12} for styles 1 and 2. The headway distribution and the relative velocity distribution are similar in form with each other as well as the corresponding distributions obtained for the real traffic, see, e.g., \cite{wagner2003empirical,wagner2006human,wagner2011time,wagner2012analyzing}. 

As far as the distribution of the car acceleration and jerk are concerned, both of them contain some sharp peaks, which requires additional analysis in order to clarify the their roles. 

\begin{figure}[p]
	\begin{center}
		subject 2, $V=80$~km/h (style 1)\\
		\hspace{0.15\textwidth}full set\hspace{0.30\textwidth} reduced set, $p> 0.25 $\\
		\includegraphics[width=1.0\textwidth]{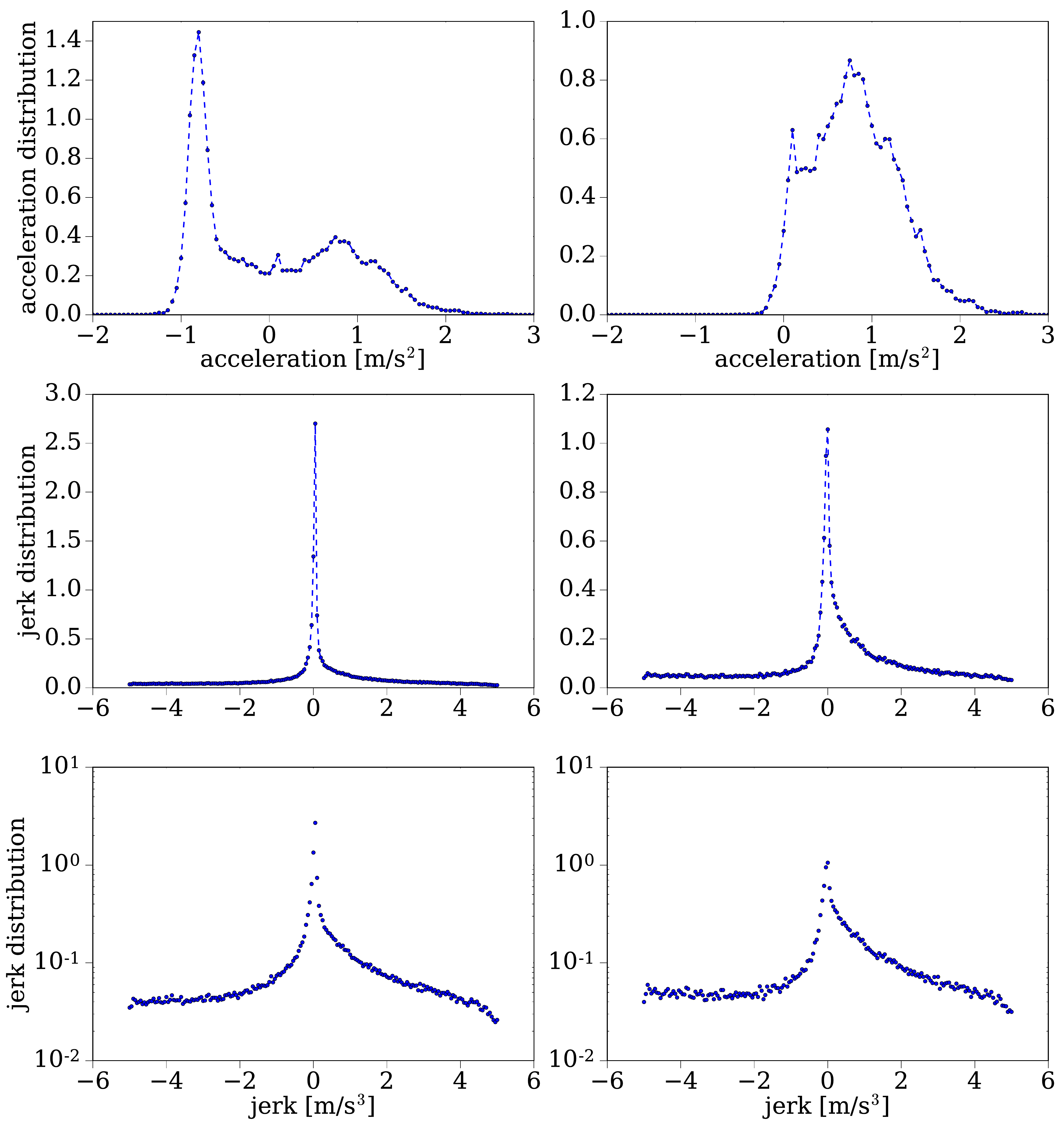}
	\end{center}
	\caption{The forms of the acceleration and jerk distributions shown in Fig.~\ref{fig:hvaj_styles12} redrawn in normal scales (upper two rows) and log-log scales (lower two rows).}
	\label{fig:aj_styles1aj}
\end{figure}

For style 1 the detailed forms of the acceleration and jerk distributions is shown in Fig.~\ref{fig:aj_styles1aj}. As seen within normal scales the acceleration distribution contains a relatively sharp peak at the acceleration $a_0\approx 1$~m/s$^2$. This peak, however, should be an apparent consequence of the driving style 1. Namely, releasing the acceleration pedal a driver fixes in some way its position precisely and the deceleration rate takes a special value. To justify this hypotheses we analyzed the statistics of subject's actions removing the points of the pedal position less then a certain threshold $p_\text{th}$. The right column in Fig.~\ref{fig:aj_styles1aj} these reduced data for $p_\text{th} = 0.25$ (the values of $p$ are given in units of the maximal pedal shift). As seen, in this case  the first peak is absent.  Therefore within style 1 the first peak is the result of releasing the acceleration pedal. The second rather smooth peak correspond to short time intervals of pushing the pedal without the accuracy required for following the lead car in the steady-state way. Contrarily, the jerk distribution possesses a sharp peak for the full data set as well as the reduced one. The latter argues for the statement that the jerk must be the direct order parameter and belong to the list of phase variables describing the car-following.

\begin{figure}[p]
	\begin{center}
			subject 7, $V=120$~km/h (style 2)\\
			acceleration\hspace{0.40\textwidth}jerk\\
		\includegraphics[width=1.0\textwidth]{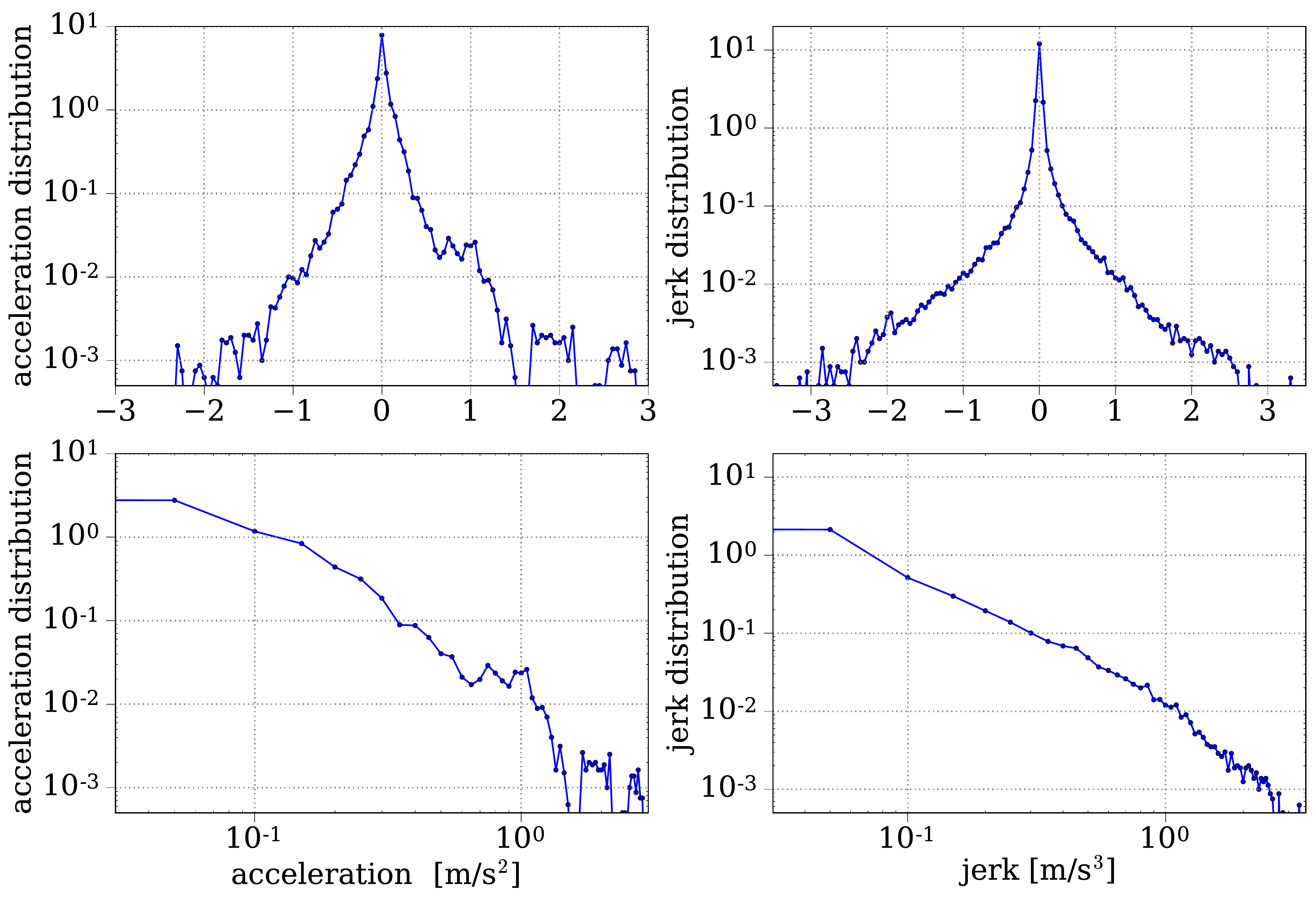}
	\end{center}
	\caption{The characteristic form of the acceleration and jerk distributions for style 2 that are represented in the log-normal scales (upper plots) and log-log scales (lower plots).}
	\label{fig:aj_styles12}
\end{figure}

For style 2 the difference between the distribution of the car acceleration and jerk is not so clear because the peak of acceleration distribution and that of the jerk distribution look rather similar; both of them are approximately of the same thickness and located at the origin. So, appealing to these plots it is difficult to recognized which variable---the acceleration or the jerk---causes the appearance of the peak of the other variable. Figure~\ref{fig:aj_styles12} depicts the acceleration and jerk distributions again in log-normal scales as well as the log-log scales. As seen, the jerk distribution is much more regular, its form much more closer to the power law
\[
\mathcal{P}(j)\approx \frac1{j^\beta}
\]
where $\beta$ is a constant. It allows us to attribute the leading role the car jerk and regard it as the phase variable of the car dynamics. 

\begin{figure}[p]
	\begin{center}
        \  subject 2, $V=80$~km/h (style 1) \qquad subject 7, $V=120$~km/h (style 2)\\
		\includegraphics[width=0.49\textwidth]{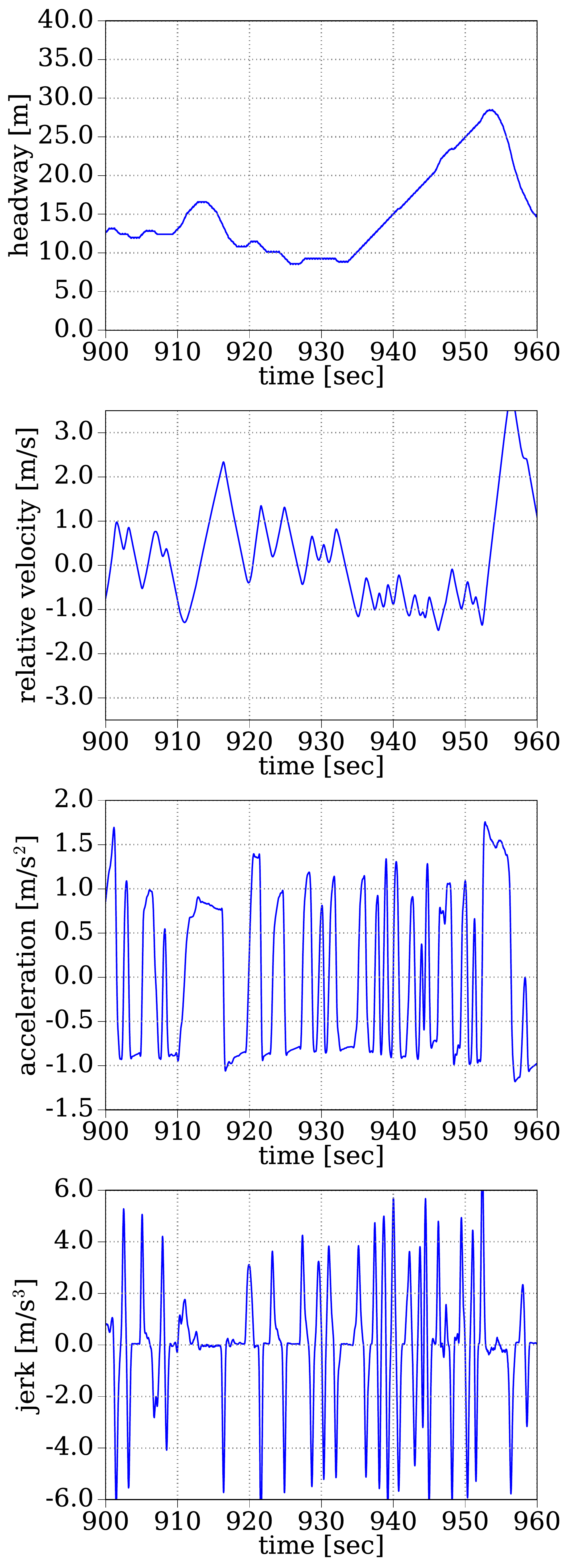}
		\includegraphics[width=0.49\textwidth]{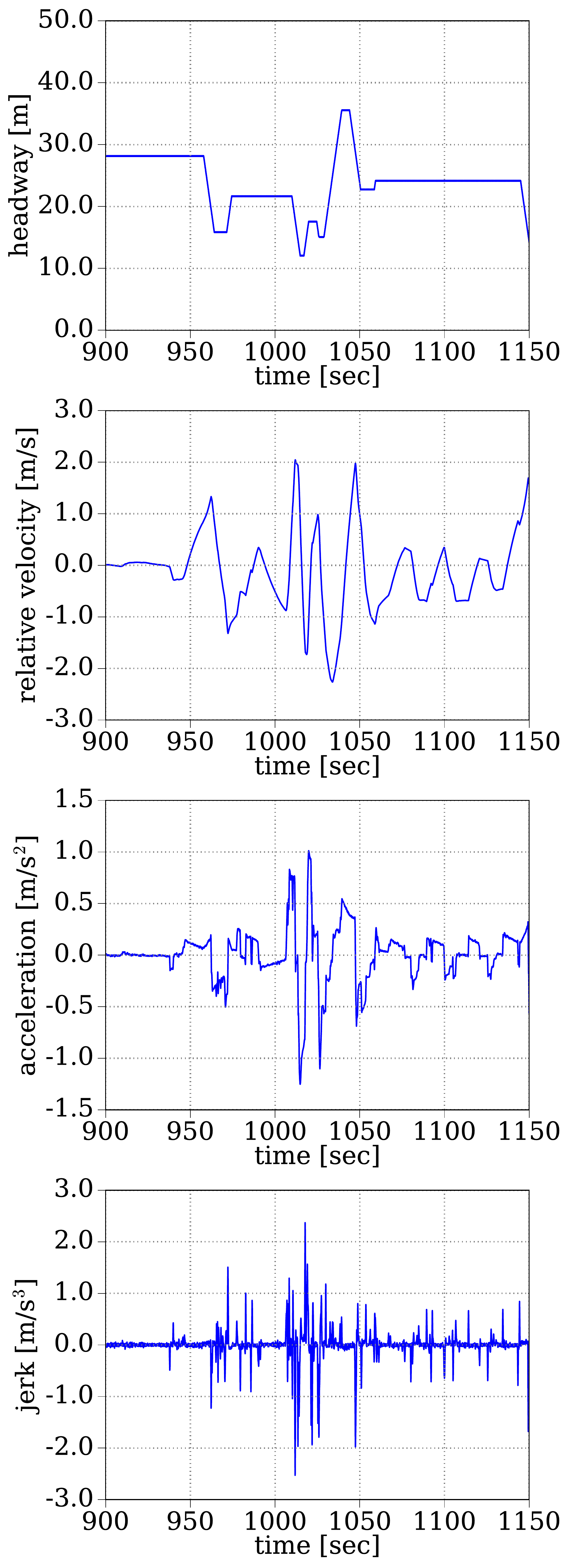}
	\end{center}
	\caption{Characteristic fragments of the time patterns of the headway, the car velocity, acceleration, and jerk found for style 1 and style 2. Duration of the shown fragments is about 1--2 minutes.}
	\label{fig:timepatterns}
\end{figure}

The drawn conclusion about the leading role of the car jerk is also justified by time patterns of the headway distance, the car velocity, acceleration and jerk shown in Fig.~\ref{fig:timepatterns}. As seen, only the jerk demonstrates the time pattern typical for the human intermittent control; they are a sequence of alternate phases of subject's passive and active behavior, where the passive phase corresponds to a certain fixed parameter  and the active phase fragments show dynamic variation of this parameter.

The last two Figure~\ref{fig:2Dcorrelations} demonstrates the relationship between the variables of the car dynamics and human actions. Namely based on the presented data for both the styles~1 and 2 we can write the relationship between the pedal position $p$, the car acceleration $a$ and the jerk 
\[
j = \frac{da}{dt}
\] 
in the form
\begin{equation}\label{maineq}
 \tau_c j = \kappa \theta - a\,, 
\end{equation}
where the time scale $\tau_c$ and the coefficient $\kappa$ should characterize the mechanical characteristics of car. The existence of the transient term---the right hand side of equation~\eqref{maineq}---becomes clear if we turn to the subject 7 data (style 2) shown in Fig.~\ref{fig:2Dcorrelations}. Subject~7 preferred to keep the acceleration pedal at a fixed position for a relatively long time and the jerk change for fixed pedal position becomes visible in the car state distribution on the plane \{car jerk--pedal position time derivative\}.  Nevertheless, it should be a minor effect and the steady-state approximation of relationship~\eqref{maineq} 
\begin{equation}\label{maineqst}
a = \kappa \theta 
\end{equation}
can be used in modeling the car-following.  

\newcommand{\myindent}{0.3\textwidth}
\begin{figure}[p]
	\begin{flushleft}
		\hspace{\myindent} subject 2, $V=80$~km/h (style 1) \\
		\hspace{0.2\textwidth} full set \hspace{0.35\textwidth} reduced set, $p > 0.25$\\
		\includegraphics[width=0.495\textwidth]{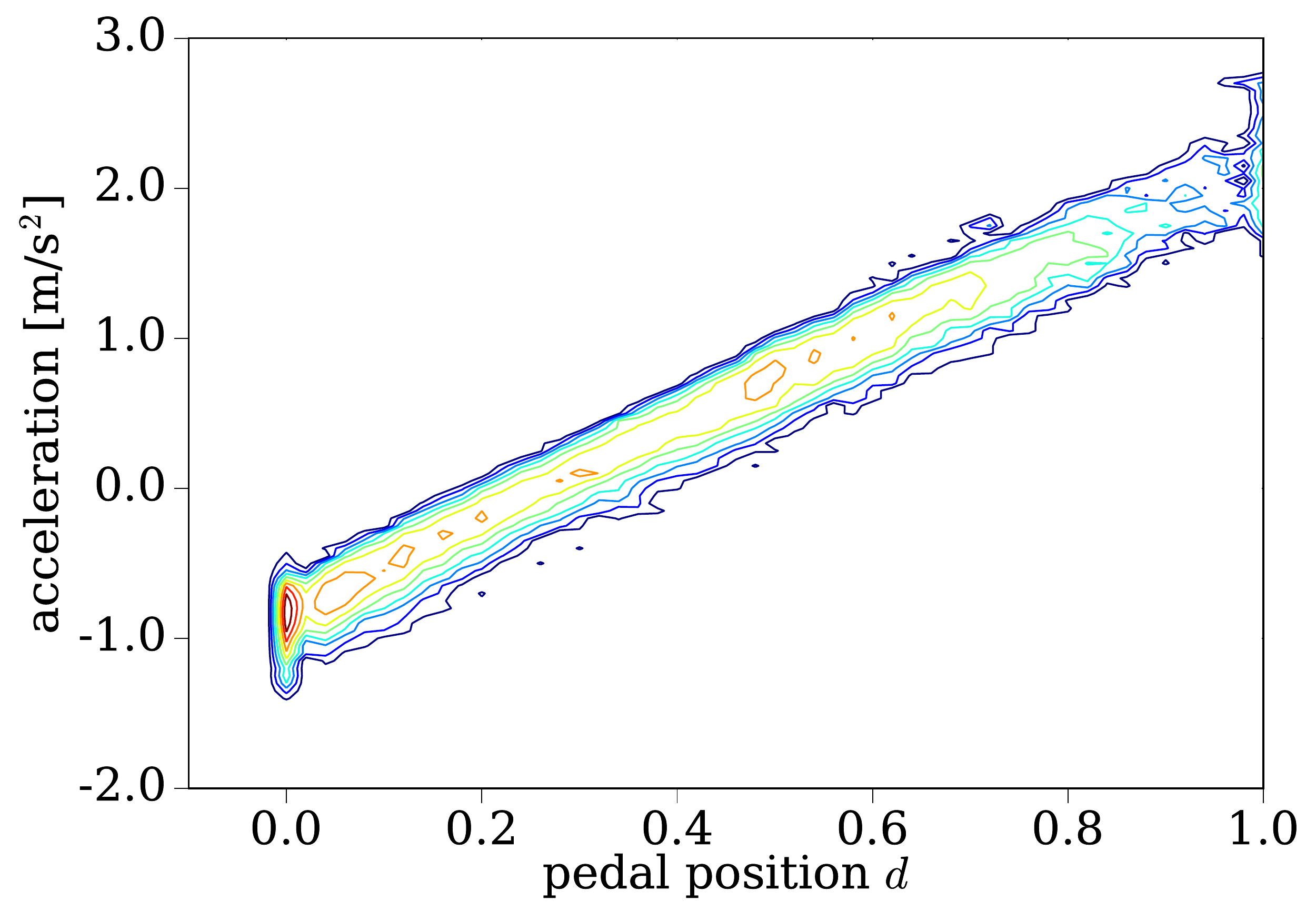}
		\includegraphics[width=0.495\textwidth]{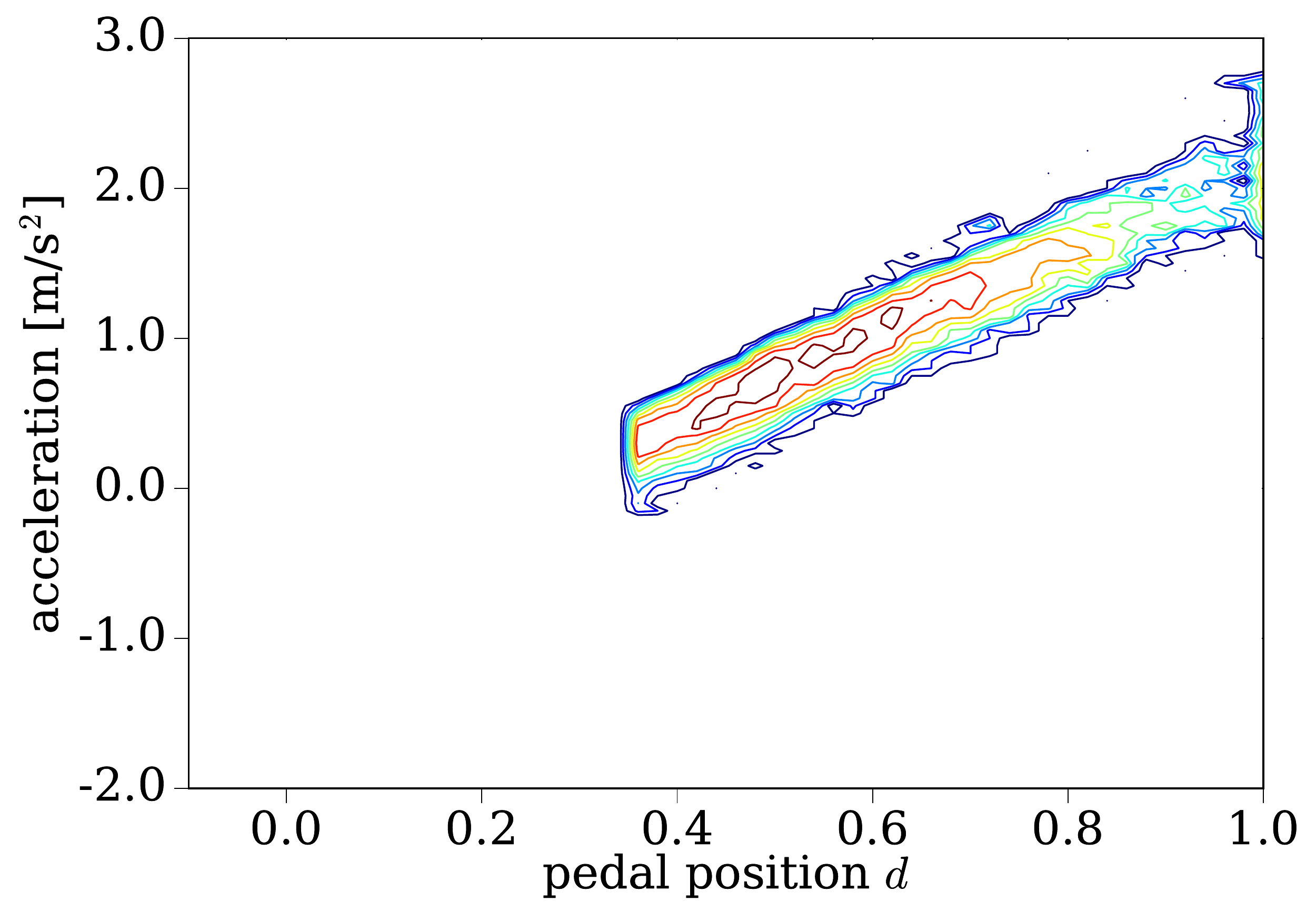}\\
		\includegraphics[width=0.485\textwidth]{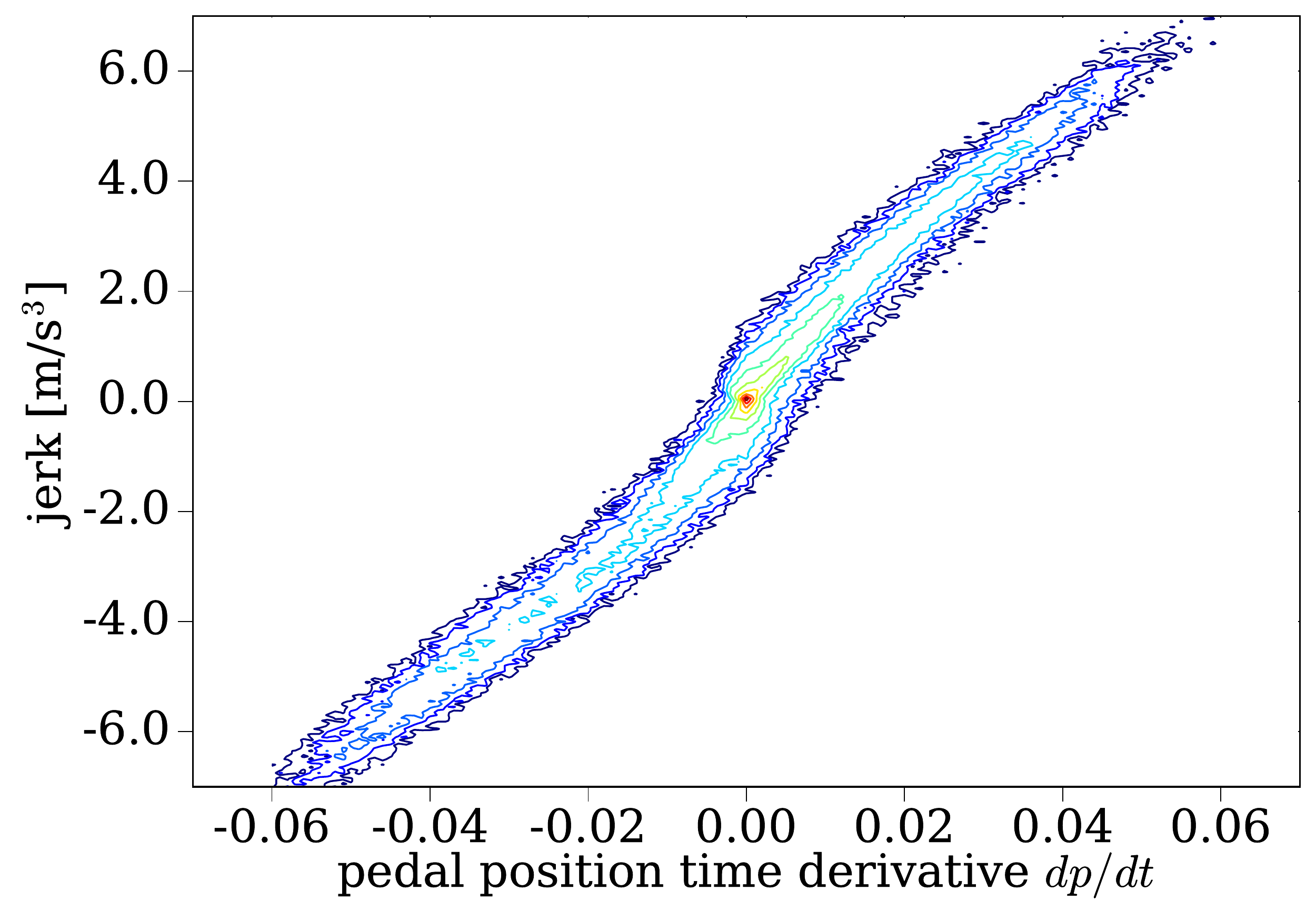}\hspace{1mm}
		\includegraphics[width=0.485\textwidth]{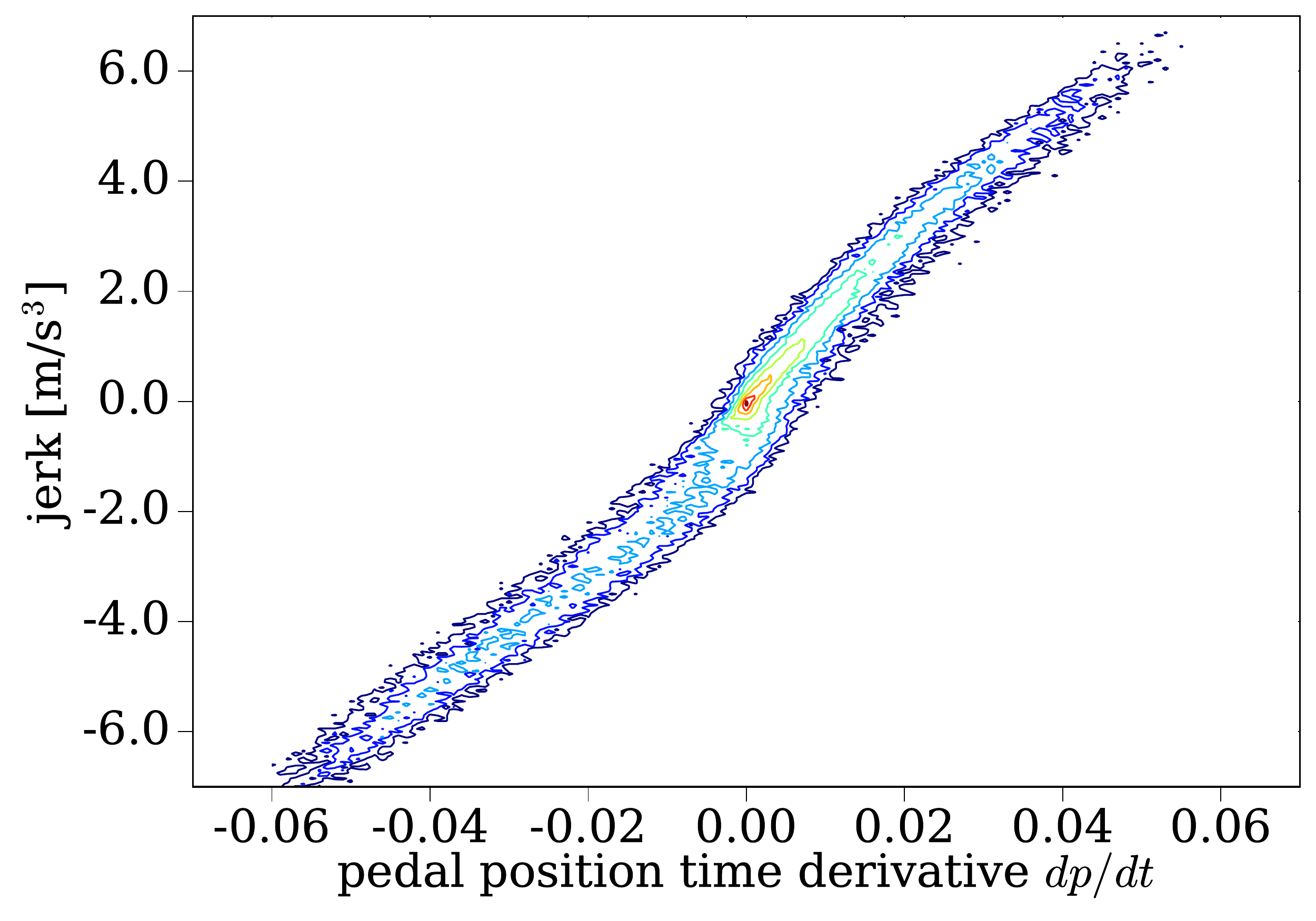}\\[1.5\baselineskip]
		\hspace{\myindent} subject 7, $V=120$~km/h (style 2, full set)  \\
		\includegraphics[width=0.495\textwidth]{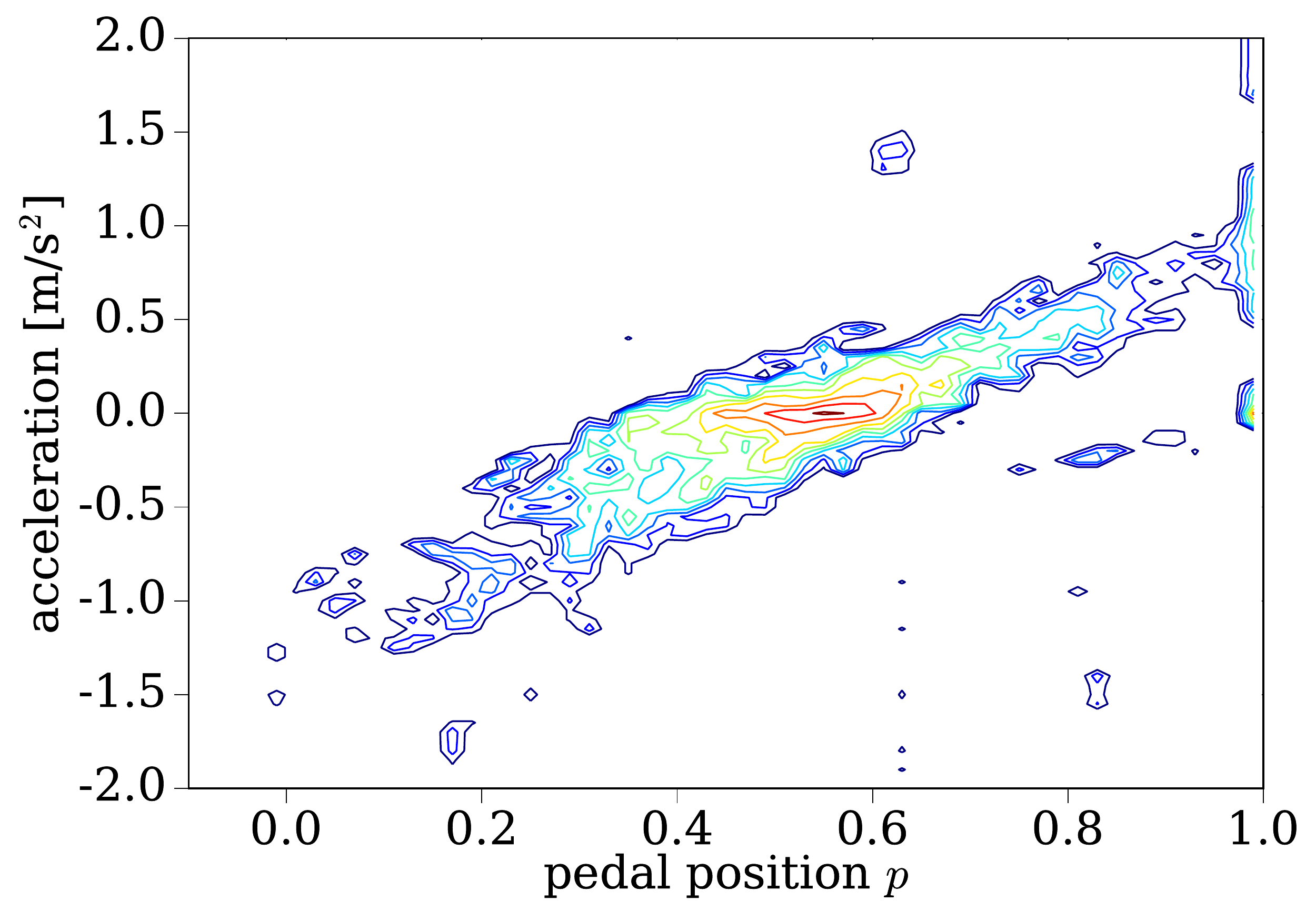}
		\includegraphics[width=0.495\textwidth]{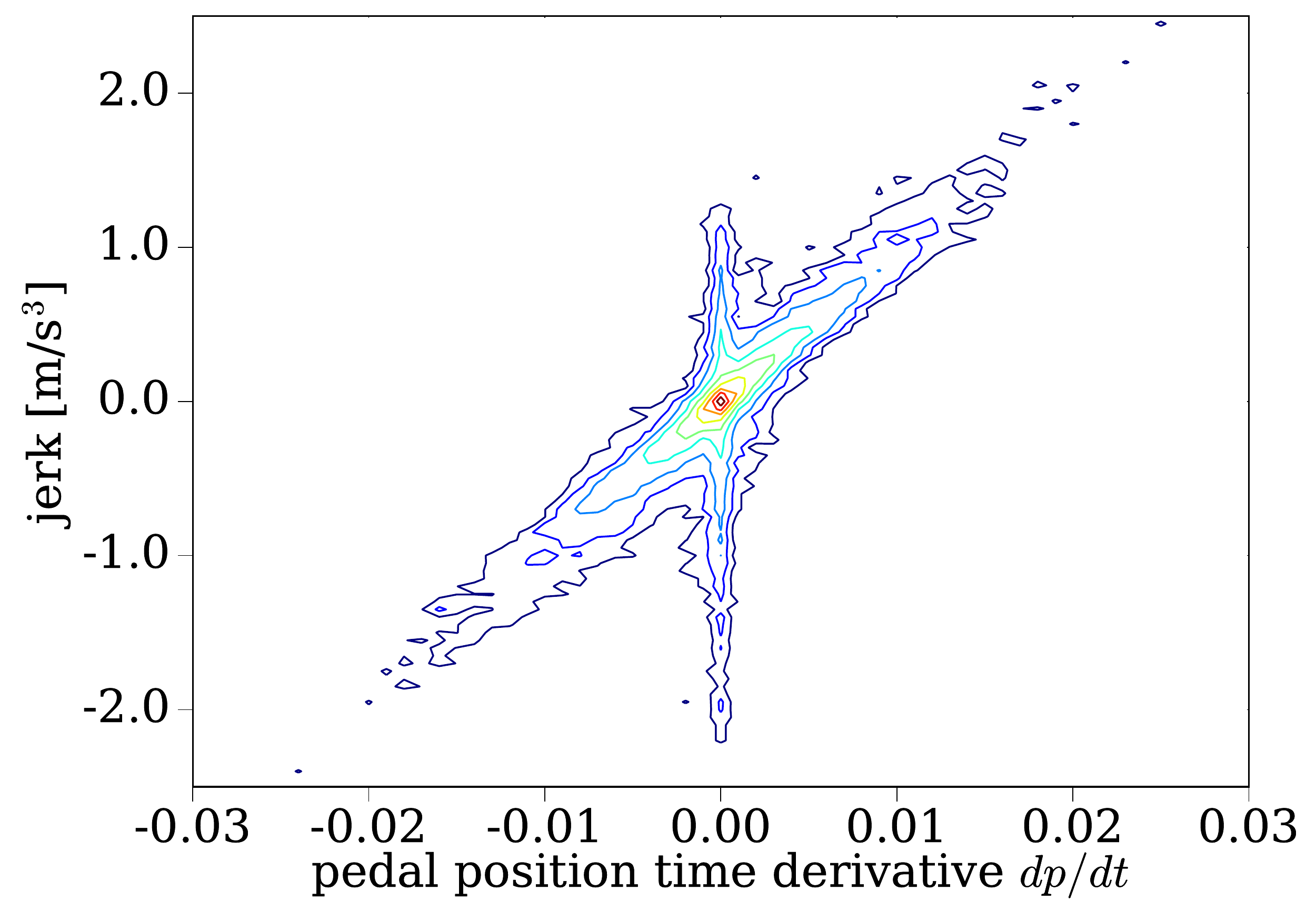}
	\end{flushleft}
	\caption{The system state distribution on the planes  \{car acceleration--pedal position\} and \{car jerk--pedal position time derivative\} for styles 1 (two upper rows) and 2 (lower row). The level lines (10 lines) are plotted in logarithmic scale.}
	\label{fig:2Dcorrelations}
\end{figure}

\newpage
\section{Four-Variable Model of Car-Following}

In this section we discuss a mathematical model for car-following that employs the results of the experiments noted above. It is based on the assumption that to describe the driver behavior the extended phase comprising four independent variables is required; this idea was partly elaborated in \cite{lubashISCIE14driving}. A driver is not able to change the car position and its velocity directly; he can only vary the car acceleration by pressing the acceleration or brake pedal. In real driving the car acceleration on its own is an important characteristic of car motion. Therefore, in describing the car dynamics we have to include the car acceleration in the list of the phase variables \cite{lubashevsky2003rational, lubashevsky2003bounded, zgonnikov2014extended}. However, according to the found characteristics of the jerk distribution in the present research as well as in our preliminary experiments  \cite{lubashISCIE14driving}, \textit{the jerk on its own is also an independent phase variable} or another additional variable combining the headway distance $h$, the car velocity $v$, acceleration $a$, and jerk $j=da/dt$ within a certain relationship should be introduced. In the proposed model using a simplified description of car motion control, this fourth variable is the position $\theta$ of an effective pedal combining the gas and break pedals into one control unit.

\begin{figure*}[t]
	\begin{center}
		\includegraphics[width=0.9\textwidth]{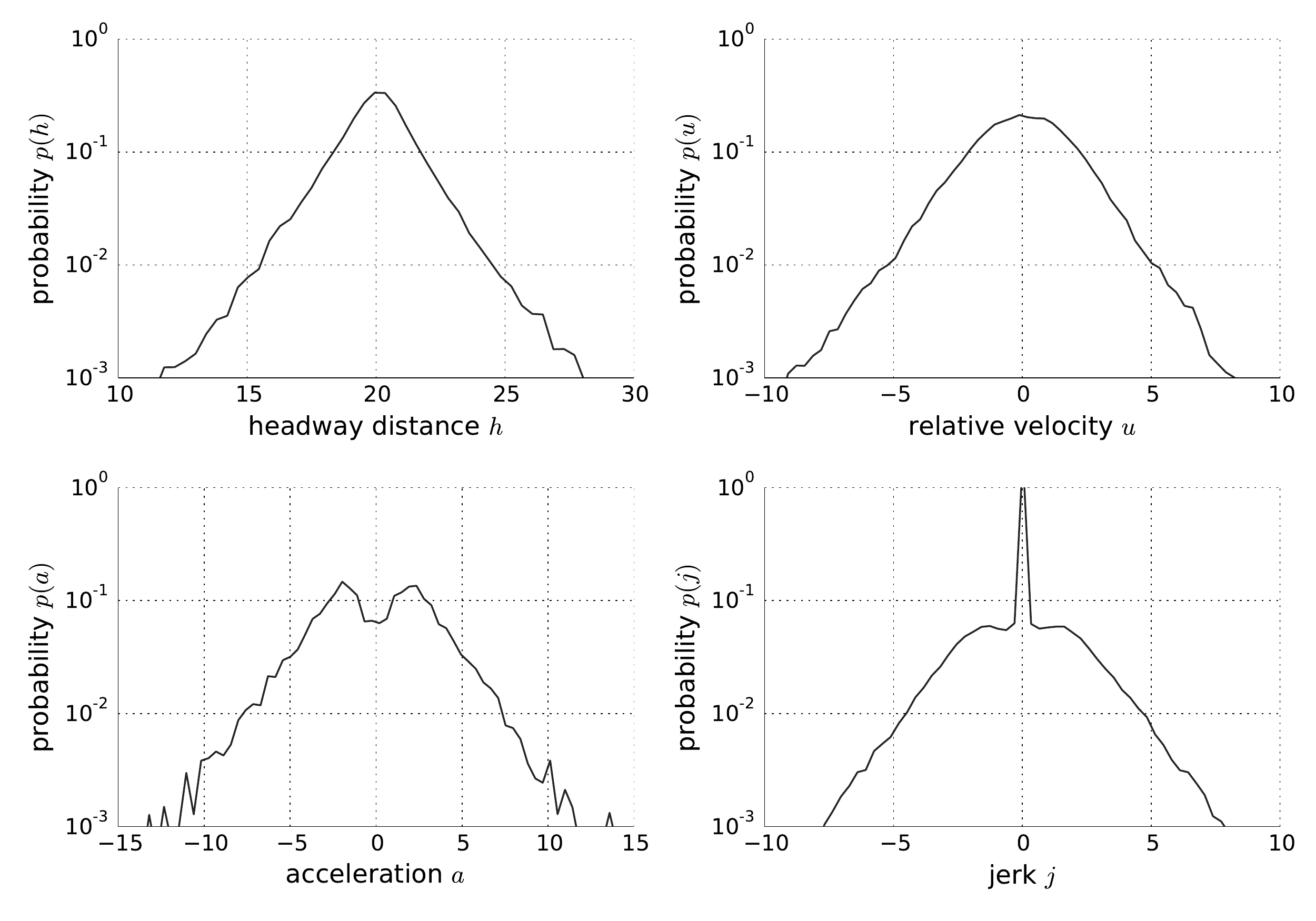}
	\end{center}
	\caption{The distributions of the headway distance $h$, relative velocity $u=v-V$, acceleration $a$, and the jerk $j\propto (\theta-v)$ obtained by numerical solution of  model~\eqref{m:1}--\eqref{m:4}. In simulation the following values were used, $v_\text{max} = 30$~m/s (about 100~km/h), $V = 15$~m/s, $D=20$~m, $a_\text{th}=0.1$~m/s$^2$, $\tau_h$ = $\tau_\theta = 0.2$~s, $\tau_v=1$~s, and $\epsilon=0.005$~m/s$^{1.5}$. The numerical labels at axes are given in the corresponding units composed of meters and seconds.}
	\label{F:M}
\end{figure*}

Namely, the model is specified as follows. The car ahead is assumed to move at a fixed velocity $V$ and the dynamics of the following car is given by the equations
\begin{eqnarray}
	\label{m:1}
	\frac{dh}{dt} & = & V-v\,,\\
	\label{m:2}
	\frac{dv}{dt} & = & a\,, \\
	\label{m:3}
	\tau_{\theta}\frac{da}{dt} & = & \theta-a\,, \\
	\label{m:4}
	\tau_{h}\frac{d\theta}{dt} & = & \Omega\left(a-\theta \right)\cdot\left[a_{\text{opt}}(h,v)-a\right] +\epsilon\xi(t)\,.
\end{eqnarray}
Equations~\eqref{m:1} and \eqref{m:2} are just simple kinematic relations between the variables $h$, $v$, and $a$, equation~\eqref{m:3} describes the mechanical properties of the car engine and its response with some delay $\tau_\theta$ to the position $\theta$ of the control unit measured here in units of acceleration. The last equation~\eqref{m:4} describes the driver behavior. It combines the basic ideas of noise-driven activation in human intermittent control \cite{zgonnikov2014Inerface} and the concept of action dynamical traps for systems with inertia \cite{zgonnikov2014extended}. The driver is able to control directly only the position $\theta$ of the control unit and the difference $(\theta - a)$ between the desired acceleration $\theta$ and the current car acceleration $a$ is the parameter quantifying the difference between his active and passive behavior. The bounded capacity of driver cognition is described in terms of action dynamical traps via the introduction of cofactor         
\begin{equation}
	\Omega(a-\theta)  = \frac{(a-\theta)^{2}}{(a-\theta)^{2}+a_{\text{th}}^{2}}
\end{equation}
similar to fuzzy reaction coefficients. Here $a_\textit{th}$ is the driver perception thres\cite{loram2011human}hold of car acceleration.  The ansatz
\begin{equation}
	a_{\text{opt}}(h,v)=\frac{1}{\tau_{v}}\left[v_{\text{max}}\frac{h^{2}}{h^{2}+D^{2}}-v\right]
\end{equation}
determines the optimal acceleration with which the strictly rational driver with perfect perception would drive the car. This expression inherits the optimal velocity mode widely used in modeling traffic flow (see, e.g.. \cite{kesting2012traffic}). Here $\tau_v$ is the human response delay time, $v_\text{max}$ is the maximal velocity acceptable for safety reasons on a given road without neighboring cars, and $D$ is the characteristic headway distance when drivers consider it necessary to slow their cars down as the headway distance decreases. The last term in equation~\eqref{m:4} is the random Langevin force, where $\xi(t)$ is white noise of unit amplitude and $\epsilon$ is the Langevin force intensity. The interplay between the fuzzy perception function $\Omega(\theta-a)$ and this Langevin force are two main components of the noise-induced activation model elaborated in \cite{zgonnikov2014Inerface} for describing human balance of overdamped pendulum. Finally, the difference $[a_{\text{opt}}(h,v)-a]$ quantifies the stimulus for the driver to correct the current state of car motion.

It should be noted that this approach to describing the effects caused by the bounded capacity of human cognition inherits the general formalism developed previously and called the dynamical traps \cite{lubashevsky1998anomalous, lubashevsky2003noised, lubashevsky2005long, lubashevsky2012dynamical}. It assumes that individuals (operators) governing the dynamics of a certain system try to follow an optimal strategy in controlling its motion but fail to do this perfectly because similar strategies are indistinguishable for them. In systems,  where the optimal dynamics implies the stability of a certain equilibrium point in the corresponding phase space, the human fuzzy rationality gives rise to some neighborhood of the equilibrium point, the region of dynamical traps, wherein each point is regarded as an equilibrium one by the operator. So, when the system enters this region and while it is located in it, maybe for a long time, the operator control is suspended. In this case the system can leave the dynamical trap region only because of the mismatch between actions which may be treated as some random factor.

In the given section we actually present a preliminary investigation of this model and its goal is to demonstrate a potential capability of such an approach to describing complex properties of real traffic flow. Figure~\ref{F:M} depicts the results of numerical solution of model~\eqref{m:1}--\eqref{m:4} using the characteristic values of the systems parameters employed by other models, at least, being of the same order (cf., e.g., \cite{kesting2012traffic}). It should be noted that the distributions obtained by numerical simulation of the developed model and constructed based on the experimental data collected by subjects with experience of driving real cars (see the present research and preliminary one \cite{lubashISCIE14driving}) look rather similar.

\newpage
\section{Conclusion}

In the last decades there has been developed is a novel concept of how human operators govern mechanical systems including unstable; it is called human intermittent control \cite{loram2011human}. This concept considers humans not to be capable of controlling system dynamics continuously. As a result, their actions must be a sequence of alternate phases of active and passive behavior, with the switching between these phases being event-driven. Recently \cite{zgonnikov2014Inerface}, a concept of noise-driven control activation was developed as a more advanced alternative to the conventional threshold-driven activation. It assumes the transition from passive to active phases to be probabilistic, which reflects human fuzzy evaluation of the current system state before making decision about the necessity of correcting the system dynamics. During the passive phase the control is suspended and the system moves on its own, broadly speaking, during the passive phase the operator accumulates the information about the system state. The periods of active phase can be regarded as fragments of open-loop control, which is due to the delay in human reaction (e.g., \cite{loram2011human}). Based on the experiments on balancing overdamped virtual pendulum \cite{zgonnikov2014Inerface} it has been demonstrated that the human action intermittency manifests itself in a sharp peak appearing at the distribution of the quantity directly controlled by the operators and this quantity has to be one of the phase variables describing the system dynamics.

Driving a car in following a lead car is a characteristic example of human control, which allows us to suppose that the intermittency of human control should be pronounced in the driver behavior and affect the motion dynamics essentially. The general objective of our study is to elucidate how the basic properties of human control manifest themselves in the characteristics of car driving. Another reason of choosing the given subject is to understand which factors are responsible for the found characteristic properties of real traffic flow, at least, some of them. 

The present thesis comprises two parts. In the first one we report the results of our experimental investigation of human actions in the car-following within a virtual environment. The used car-driving simulator was created based on the available open source engine TORCS and employed for analyzing subject's behavior in the car-driving within the car-following setup. A commercially available high-precision steering wheel and the pedal set (Logitech G27 Racing Wheel) was used. A special track of rectangular form (with smoothed corners) where the longest straight road part is of 70 km was created. The width of this road is 15 m and its roadside includes a special pattern of stripes enabling a subject to get feeling of the current speed. Eight subjects with different skill in driving real cars participated in these experiments. They were instructed to drive a virtual car without overtaking the lead car driven by computer at a fixed speed and not to lose sight of it. In the experiments four different lead car speeds, 60km/h, 80km/h, 100km/h, and 120km/h, were studied. The parameters of the car model were set such that the virtual car dynamics be similar to the dynamics of real cars belonging to the intermediated class between the subcompact and compact cars.
Based on the collected data the statistical properties of the headway distance, the car velocity, acceleration, and jerk as well as the pedal position and its time derivative are studied in detail. It is demonstrated that the pedal position time derivative and the related variable---the car jerk---are actually the quantities via which a driver governs the car dynamics. As found, the distributions of these variables always possess sharp peaks at the origin determining the properties of possible peaks of other quantities, e.g., the car acceleration. In particular, it means that the driver actions should be categorized as the intermittent control and the car jerk (related directly to the time derivative of the pedal position) has to be included in the list of the phase variables describing the car-following. In other words,
the car-following problem does not meet the paradigm of Newtonian mechanics and has to be categorized as higher order time derivative problem. Besides, we singled out several styles of driving, one of the limit cases corresponds to the driver behavior when he presses and release the acceleration pedal rather often. In the opposite limit case a driver mainly keeps the pedal pressed. Indeterminate styles mix the two limit styles, which admits interpretation as mesolevel intermittency in human actions.

The second part of the thesis develops a mathematical model for the car-following that operates with the extended phase space comprising the headway distance, the car velocity, acceleration, and jerk. The jerk is regarded as the main parameter via which the control of car dynamics is implemented. The control activation is assumed to be governed by the noise-driven mechanism. Results of experiments, numerical simulation are compared with each other as well as the available data of real traffic flow. 

\newpage

\bibliography{library,Physics,TrafficPhysics}

\end{document}